\documentclass[12pt]{article}
\usepackage[utf8]{inputenc}
\usepackage{amsmath}
\usepackage{amsfonts}
\usepackage{amssymb}
\usepackage{amsthm}

\usepackage{amsmath,graphicx}
\graphicspath{{figures/}} 
\usepackage{subfigure}
\usepackage[english]{babel} 
\usepackage{graphicx,amssymb,amstext,amsmath}
\usepackage{float}
\usepackage{amsfonts}
\usepackage{bbm}
\usepackage{bm}
\usepackage{verbatim}
\usepackage[usenames]{color}			
\usepackage{flushend}
\usepackage{multirow}
\usepackage{makecell}
\usepackage[toc,page]{appendix}
\usepackage[numbers]{natbib}
\usepackage{url}
\usepackage{setspace}

\usepackage{tabu}

\theoremstyle{definition}

\newcommand{\x}{{\bf x}}
\newcommand{\y}{{\bf y}}

\newcommand{\post}{{\bar{\pi}}}

\date{ }

\title{
MCMC-driven  importance samplers}

\author{F. Llorente$^+$, E. Curbelo$^+$, L. Martino$^\dagger$*, V. Elvira$^{\ddagger}$, D. Delgado$^+$ \\
{\small$^+$ Universidad Carlos III de Madrid (UC3M), Spain}\\
{\small$^\dagger$ Universidad Rey Juan Carlos (URJC), Spain}\\
{\small$^{\ddagger}$  University of Edinburgh, UK} \\
{\scriptsize * corresponding author: \url{luca.martino@urjc.es}}
}

\begin{document}
\maketitle
\begin{abstract}
 Monte Carlo {sampling} methods are the standard procedure for approximating complicated integrals of multidimensional posterior distributions in Bayesian inference. In this work, we focus on the class  of Layered Adaptive Importance Sampling (LAIS) scheme, which is a family of adaptive importance samplers where Markov chain Monte Carlo algorithms are employed to {\it drive} an underlying multiple importance sampling scheme. The modular nature of LAIS allows for different possible implementations, yielding a variety of different performance and computational costs. In this work, we propose different enhancements of the classical LAIS setting in order to increase the efficiency and reduce the computational cost, of both upper and lower layers. The different  variants address computational challenges arising in real-world applications, for instance with highly concentrated posterior distributions.  Furthermore, we introduce different strategies for designing cheaper schemes, for instance, recycling samples generated in the upper layer  and using them in the final estimators in the lower layer. {Different numerical experiments, considering several challenging scenarios, show the benefits of the proposed schemes comparing with benchmark methods presented in the literature.}

\end{abstract}


\section{Introduction}


Bayesian methods have become very popular in statistics, signal processing, and machine learning during the last years and, with them,  Monte Carlo (MC) techniques\footnote{{In this work, with MC techniques, we refer to the well-known MC sampling methods formed by the following four main families: (a) direct methods based on transformation of random variables, (b) rejection sampling procedures, (c) Markov chain Monte Carlo (MCMC)  algorithms and (d) importance sampling (IS) schemes   \cite{Robert04}.} } that are often necessary for the implementation of optimal a-posteriori estimators  \cite{Doucet01b,Liu04b,Robert04}. {Indeed, MC methods are powerful tools for approximating integrals involving a complicated posterior distribution $\bar{\pi}(\x)=\bar{\pi}(\x|\y)$  \cite{Djuric11,Fitzgerald01,hong2010joint,Moral06}. }
\newline 
Markov chain Monte Carlo (MCMC) and importance sampling (IS) algorithms are well-known families of MC methods. They can also consider stochastic quadrature techniques  \cite{Burden00,JohariRaro,JohariRaro2,JohariRaro3}. Since both families have their own drawbacks and benefits, there have been attempts to combine them in order to design efficient schemes \cite{beaujean2013initializing,Botev13,LIESENFELD2008272,rudolf2020metropolis,schuster2020markov}.
 The general framework called {\it Layered Adaptive} IS (LAIS)  is one of such attempts  \cite{martino2016layered}, combining the desirable exploratory behavior of MCMC, and the robustness (and easier theoretical validation) of IS. The main underlying idea of this algorithm is the {\it layered} (i.e. hierarchical) procedure for generating samples. In order to generate one sample, a location parameter is drawn from a probability density function (pdf) $\bm{\mu}_i\sim p(\bm{\mu})$ (that plays the role of a prior pdf over a location parameter in the hierarchical procedure) and, conditionally on it, a sample is generated from a proposal density centered at $\bm{\mu}_i$, i.e.,  $\x_i \sim q_i(\x|\bm{\mu}_i)$. Then, the sample $\x_i$ is properly weighted according to a {\it multiple IS} (MIS) procedure \cite{elvira2015generalized,Veach95}.
Hence,  the {\it upper layer} is formed by the generation of $\bm{\mu}$'s, while in the {\it lower layer}, we have the generation of  $\x$'s and its weighting.  More generally,  {parallel MCMC algorithms} addressing different $p_i(\bm{\mu})$'s, for $i=1,...,N$, can be employed to obtain the  location parameters $\bm{\mu}_i$. {The use of parallel MCMC chains  in the upper layer makes LAIS particularly suitable in multimodal scenarios.
 Note that the samples $\bm{\mu}_i$ are not included in the final estimators (as the samples $\x_i$), but only used as location parameters for the proposal densities.  In \cite{martino2016layered}, the specific choice $p_i(\bm{\mu})=\bar{\pi}(\bm{\mu})$ has been suggested and successfully tested.   }
\newline
\newline
 { 
 With respect to other benchmark AIS techniques in the literature \cite{Cappe04,Cornuet12,ELVIRA201777}, LAIS provides very competitive results and exhibits a relevant robustness with respect to tuning of the parameters of proposal densities $q_i$ (such as the scale parameters). The interested reader can observe these properties in the numerical comparison, provided in Section \ref{Robustness}. Moreover, LAIS can be interpreted as:
 \begin{itemize}
\item An efficient procedure of combining the outputs of several parallel MCMC chains. Several other attempts can be found in the literature \cite{Calderhead14,Corander08,Corander06,earl2005parallel,Jacob11}.
\item An efficient procedure for estimating the marginal likelihood by using MCMC chains (which is a well-known difficult task for the MCMC techniques \cite{llorente2020marginal}).
\end{itemize}
 These strength points of LAIS are very appealing for practitioners and researchers.} 
At the same time, the generic LAIS framework offers {a remarkable flexibility}  
which have not been completely exploited in \cite{martino2016layered}, {and have been not explored in the further works.}
For instance, in the upper layer, the user must specify the choices of $p_i(\bm{\mu})$ and the type of MCMC algorithms; in the lower layer, a specific MIS weighting scheme must be selected. 
 { This flexibility allows LAIS to handle efficiently different complex inference scenarios, not only the multimodality. Introducing specific LAIS schemes for tackling other difficult scenarios of inference is the first main goal of this work. The second main objective of this paper is to describe different procedures for reducing the computational cost of LAIS. }
 \newline
  \newline
In this work, as disclosed above, we introduce different schemes for improving the overall performance and reduce the total computational cost of LAIS.  { Specifically, we discuss suitable configurations of LAIS for addressing the problem of sampling concentrated posteriors (due to complex model or the great number of data) and posteriors in high dimensional spaces. This is possible by the use of data-tempered posteriors in the upper layer, that we refer to as {\it partial posteriors} (see Section \ref{DataTempering}), and advanced MCMC schemes such Hamiltonian MC (HMC) and sophisticated Gibbs-type techniques \cite{Neal11,Robert04,Gilks92} (see Section \ref{UppAS}).}  We also discuss different strategies for reducing the overall computational cost. For instance, we propose a procedure for recycling the samples in upper layer and use them in the final estimators, in such a way that the sampling step in the lower layer can be avoided. This drastically reduces the number of evaluations of the posterior. 
Moreover, in the lower layer, the cost of weighting can be quite high if we have run  long MCMC chains in the upper layer.
 This problem can also be alleviated by using ideas such as compression or alternative weighting schemes, that reduce the cost but maintain the same performance for the final estimators \cite{el2019efficient,martino2021compressed}.
 We test the variants in different scenarios with synthetic and real data.\footnote{Related Python and Matlab codes are available at \url{https://github.com/FLlorente/LAIS_extensions.}}  A theoretical discussion about the optimal choice of $p({\bm \mu})$ is also provided in the Appendix. {Several numerical simulations show  the benefits of the proposed LAIS techniques in different challenging sampling problems. Table \ref{TableOFcontribution} summarizes the main contributions (and the novel schemes) and main acronyms employed in this work. Finally, Table \ref{TableNotation} summarizes the main notation of the work.\footnote{In Table  \ref{TableNotation}, with the acronym MH, we denote the Metropolis-Hastings algorithm \cite{Robert04}.}}








\begin{table}[!h]
	\centering
		{
	\caption{Summary of the main contributions and the main acronyms of the work.} \label{TableOFcontribution}
	\vspace{0.1cm}
	\footnotesize
	\begin{tabular}{|c|c|c|c|}
		\hline 
		{\bf Contribution/Proposed scheme} & {\bf Section}  & {\bf Reducing cost} & {\bf Improving performance}  \\
		\hline
		 Partial posteriors LAIS (PLAIS) & \ref{DataTempering} &   \checkmark & \checkmark    \\
		 Hamiltonian-driven IS (HMC-LAIS)   & \ref{UppAS} & &   \checkmark  \\
		  Gibbs-driven IS (Gibbs-LAIS) & \ref{UppAS}& &  \checkmark \\
		Compressed LAIS (CLAIS)  & \ref{sec_compression} & \checkmark &  \\
		Recycling LAIS (RLAIS)  & \ref{RLAISsect}& \checkmark & \\
		 Partial posteriors RLAIS (PA-RLAIS) & \ref{RLAISsect} &\checkmark& \checkmark  \\
		 \hline
		 \hline
		  Discussion about the computation cost  &  \ref{sec_compt_costs}  & related  & \\
		  Numerical comparisons  & \ref{sec_experiments}   &  related   &  related  \\
		 Theoretical discussion & App. \ref{Choiceofpmu}-\ref{HI_MH} & &   related  \\
		\hline
	\end{tabular}
	}
\end{table}

\begin{table}[!h]
	\centering
		{
	\caption{Main notation of work.} \label{TableNotation}
	\vspace{0.1cm}
	\footnotesize
	\begin{tabular}{|c|l||c|l|}
		\hline 
	${\bf x}\in \mathcal{X}_{\texttt{tot}}\subseteq \mathbb{R}^{D_X}$ & vector of parameters to infer & ${\bar \pi}(\x|\y_\texttt{tot})$ & normalized full posterior \\
		${\bf y}_{\texttt{tot}}$ & data &    $\pi(\x|\y_\texttt{tot})$ & unnormalized full posterior  \\
		${\bf y}_{n}$  & subset of data & $\bar{\pi}_n({\bf x}|{\bf y}_n)$ & normalized partial posterior \\
		$D_Y$ & total number of data (in ${\bf y}_{\texttt{tot}}$) & $\pi_n({\bf x}|{\bf y}_n)$ & unnormalized partial posterior\\
		$K_n$&  number of data in ${\bf y}_{n}$  &  $L({\bf y}|{\bf x})$ & likelihood function\\
		$q_{n}(\x|\bm{\mu}_{n})$  & proposal density in the lower layer & $g(\x)$ & prior density  \\
		$\varphi_{n}(\x|\bm{\mu}_{n})$  & proposal density within MH (in RLAIS) & $Z=p({\bf y}_{\texttt{tot}})$ & marginal likelihood \\
		$\bm{\mu}_{n}$ & location parameter (e.g., mean) & {\bf I} & integral of interest \\
		$N$ & number of the MCMC chains &  $w=\frac{\pi(\x|{\bf y}_{\texttt{tot}})}{{\bm \Phi}(\x)}$ & importance weight \\
		$T$ & length of  the MCMC chains & ${\bm \Phi}(\x)$ & denominator in MIS weights\\
		$M$ & number of samples per proposal  & ${\bm \Psi}(\x)$ & denominator in MIS weights (in RLAIS) \\
		$B$ & number of sub-regions $\mathcal{X}_m$  & $\mathcal{X}_m$ & $m$-th sub-region,  $\mathcal{X}_{1} \cup \ldots \cup \mathcal{X}_{B}=\mathcal{X}_{\texttt{tot}}$\\
		\hline
	\end{tabular}
	}
\end{table}

\section{Problem statement}\label{ProblemStatement}


We are interested in making inference about the vector ${\bf x}=[ x_{1},\ldots, x_{D_X}] \in\mathcal{X}_{\texttt{tot}}\subseteq\mathbb{R}^{D_X}$. We receive a set of $D_Y$ measurements, ${\bf y}_{\texttt{tot}}=[y_1,y_2,\dots,y_{D_Y}]$, 
with each $y_j\in \mathbb{R}$, related to the variable of interest ${\bf x}$.\footnote{We assume $y_j$ to be scalar only for the sake of simplicity.}
We denote the complete likelihood function as $L({\bf y}_{\texttt{tot}}|{\bf x})$. 
Considering a prior probability density function (pdf) $g({\bf x})$, the {\it complete posterior} pdf can written as
\begin{align}
\bar{\pi}({\bf x}|{\bf y}_{\texttt{tot}})=\frac{1}{p({\bf y}_{\texttt{tot}})}  L({\bf y}_{\texttt{tot}}|{\bf x})g({\bf x})=\frac{1}{Z} \pi({\bf x}|{\bf y}_{\texttt{tot}}),
\end{align}
where we have denoted $Z=p({\bf y}_{\texttt{tot}})$,  and $\pi({\bf x}|{\bf y}_{\texttt{tot}})=L({\bf y}_{\texttt{tot}}|{\bf x})g({\bf x})$.  { Note that $\bar{\pi}({\bf x}|{\bf y}_{\texttt{tot}}) \propto \pi({\bf x}|{\bf y}_{\texttt{tot}})$.}
\newline
\newline
\noindent
 {\bf Goal.} The objective is to make inference about the variable ${\bf x}$ given the information provided by knowledge of ${\bf y}_{\texttt{tot}}$. Generally, it s required to compute integrals of type
\begin{eqnarray}
\label{eq:IntOfInter}
{\bf I}&=&\int_{\mathcal{X}_{\texttt{tot}}} {\bf f}({\bf x}) {\bar \pi}({\bf x}|{\bf y}_{\texttt{tot}}) d{\bf x},
\end{eqnarray}
where ${\bf f}({\bf x}): \mathbb{R}^{D_X}\rightarrow \mathbb{R}^{s}$ and ${\bf I}\in\mathbb{R}^{s}$ with $s\geq 1$. { When ${\bf f}({\bf x})={\bf x}$, the integral ${\bf I}$ represents the {\it minimum mean square error} (MMSE) estimator of ${\bf x}$ \cite{Liu04b,Robert04}. } Moreover, we are also interested in the so-called {\it marginal likelihood},
\begin{align}\label{eq:IntOfZ}
	Z = p({\bf y}_{\texttt{tot}})=\int_{\mathcal{X}_{\texttt{tot}}} \pi(\x|{\bf y}_{\texttt{tot}})d\x.
\end{align}
{This quantity is particularly useful for the model selection purposes \cite{llorente2020marginal,Robert04}.}
Generally, we are not able to calculate analytically the integrals above. 
Importance sampling (IS) and Markov chain Monte Carlo (MCMC) are popular Monte Carlo techniques for approximating integrals as in Eq. \eqref{eq:IntOfInter} using random samples \cite{Liu04b,Robert04}. 
IS provides also an estimator of Eq. \eqref{eq:IntOfZ}, something that is not straightforward with MCMC (see e.g. \cite{llorente2020marginal} for a review of methods for estimating $Z$). 
In this work, we consider the  LAIS framework which mixes the benefits of MCMC and IS algorithms \cite{martino2016layered}.

\section{Layered adaptive importance sampling (LAIS)}\label{LAIS_sect}

%
LAIS is an adaptive IS framework that consists of two sampling layers, which is detailed in Table \ref{tab:LAIS_algorithm} and described next.
Let $\{q_{n,0}(\x|\bm{\mu}_{n,0})\}_{n=1}^N$  denote  an initial set of $N$ parametric proposals. In the upper layer, the location parameters of the proposals are updated  by means of MCMC algorithms.
In the simplest case, at iteration $t$, each $\bm{\mu}_{n,t-1}$ independently evolves to $\bm{\mu}_{n,t}$ ($n=1,\dots,N$) by running one iteration of a MCMC algorithm. 
More generally, the whole population $\{\bm{\mu}_{n,t-1}\}_{n=1}^N$ can be updated to $\{\bm{\mu}_{n,t}\}_{n=1}^N$, e.g., considering more sophisticated population MCMC algorithms \cite{Liu04,Liang10}.
Then, after performing $T$ such iterations, in the lower layer we sample $\x_{n,t}\sim q_{n,t}(\x|\bm{\mu}_{n,t})$ for $n=1,\dots,N$ and $t=1,\dots,T$, and assign weights for each sample (see Table \ref{tab:denoLAIS} for different weighting schemes).
The {\it layered} in LAIS amounts to the fact that the adaptation ({\it upper layer}) is independent from the sampling and weighting steps ({\it lower layer}). As an example, we can run first, e.g., $N$ parallel chains for $T$ iterations each in order to obtain the $NT$ locations parameters $\{\bm{\mu}_{n,t}\}$, and then perform standard IS with the $NT$ proposals. The  weighting procedure is done according to the so-called {\it deterministic mixture} approach \cite{elvira2015generalized,Veach95}. Some possible choices of the denominator of the importance weights are given in Table \ref{tab:denoLAIS}. Clearly, in the special case of a unique chain $N=1$,  the spatial denominator becomes  the standard IS denominator. If $N$ single MCMC steps are performed, i.e., $T=1$, then the temporal denominator becomes  the standard IS denominator.
\newline
\newline
The estimators of Eq. \eqref{eq:IntOfInter} and Eq. \eqref{eq:IntOfZ} are then given by
\begin{eqnarray}
	{\widehat {\bf I}}&=&\frac{1}{NT\widehat{Z}} \sum_{n=1}^N\sum_{t=1}^T w_{n,t}  {\bf f}({\bf x}_{n,t}), \label{eq:I_EST} \\
	\widehat{Z}&=&\frac{1}{NT}\sum_{t=1}^T\sum_{n=1}^N w_{n,t}.\label{eq:Z_EST}
\end{eqnarray} 
{Some bounds and theoretical results related to these estimators can be found \cite{Deniz2021ConvAIS}.}

\begin{table}[h]
		\caption{LAIS algorithm}
	\begin{tabular}{|p{0.95\columnwidth}|}
		\hline
		\small
		\\
		Choose $\{q_{n,0}\}_{n=1}^N$, $\{{\bm \mu}_{n,0} \}_{n=1}^N$ and the MCMC algorithms in the upper layer.	
		\newline
		\newline	
		\textbf{Upper layer (MCMC).} 
		\begin{itemize}
			\item {\bf Adaptation:}  Apply MCMC transitions with invariant pdf $p_n({\bm \mu})$, e.g., $p_n({\bm \mu})=\post({\bm \mu}|{\bf y}_{\texttt{tot}})$, i.e., 
			$$\{{\bm \mu}_{n,t-1} \}_{n=1}^N  \xrightarrow{\text{MCMC}}\{{\bm \mu}_{n,t} \}_{n=1}^N, \quad\forall   t=1,...,T. $$	
			\vspace{-0.8cm}	
		\end{itemize}
		
		\textbf{Lower layer (IS).}	
		\begin{itemize}
			\item {\bf Sampling:} $\x_{n,t}\sim q_{n,t}(\x|\bm{\mu}_{n,t})$, for all $n, t$. 
			\item {\bf Weighting:}
			\begin{align}
				w_{n,t} = \frac{\pi(\x_{n,t}|{\bf y}_{\texttt{tot}})}{{\bm \Phi}(\x_{n,t})}, \qquad \forall n,t,
			\end{align}
			where different denominators,  ${\bm \Phi}(\x_{n,t})$, are possible. See Table \ref{tab:denoLAIS}.
		\end{itemize}\\
		\hline
	\end{tabular}
\label{tab:LAIS_algorithm}
\end{table}


\begin{table}[!t]
	\centering
	\caption{Possible denominators $\bm{\Phi}(\x_{n,t})$.}
	\begin{tabular}{|c|c|c|c|}
		\hline 
		{\bf complete} & {\bf temporal}  & {\bf spatial}  & {\bf standard}  \\
		\hline
		& & & \\ 
		$\frac{1}{NT}\sum_{\tau=1}^T\sum_{i=1}^N q_{i,\tau}({\bf x}_{n,t}|{\bm \mu}_{i,\tau})$    & $\frac{1}{T}\sum_{\tau=1}^T q_{n,\tau}(\x_{n,t}|{\bm \mu}_{n,\tau})$ & $\frac{1}{N}\sum_{i=1}^N q_{i,t}(\x_{n,t}|{\bm \mu}_{i,t})$ &  $q_{n,t}({\bf x}_{n,t}|{\bm \mu}_{n,t})$   \\
		& & &  \\
		\hline
	\end{tabular}
	\label{tab:denoLAIS}
\end{table}



%

\noindent
{\bf Consistency.} LAIS can be interpreted as a standard, static IS scheme with $NT$ proposals, and the consistency only depends on the proper choice of the denominator $\bm{\Phi}({\bf x})$ in the importance weights.  In Table \ref{tab:denoLAIS}, some proper choices are provided which follows the deterministic mixture approach \cite{elvira2015generalized,Veach95}. It is important to remark that the consistency does not depend on the choice of the densities $p_n({\bm \mu})$ in the upper layer but, clearly, the efficiency of LAIS is affected by the selected pdfs $p_n({\bm \mu})$.
\newline
\newline
{
{\bf Remark.} For the sake of simplicity, we have assumed to draw only one sample $\x_{n,t}$ from each proposal $q_{n,t}(\x|\bm{\mu}_{n,t})$, in the lower layer. More generally, one could draw $M>1$ samples, $\x_{n,t}^{(1)},...,\x_{n,t}^{(M)}$ from each $q_{n,t}(\x|\bm{\mu}_{n,t})$. This is often necessary for performing a fair comparison with other AIS techniques and is an additional degree of freedom offered by the LAIS framework (see  Section \ref{Robustness} and \ref{HighDimEx}).  
However, for simplicity, in the rest of work we consider $M=1$, if it is not explicitly referred to the case $M>1$. }
\newline
\newline
{\bf  Evaluations of the posterior.} In the standard LAIS implementation (i.e. setting $p_n(\bm{\mu})=\post(\bm{\mu}|{\bf y}_{\texttt{tot}})$ for all $n$), the total number of evaluations $E$ of the posterior is $E=2NT$ { (or, more generally, $E=NT+MNT$)}, where $NT$ evaluations are performed in the upper layer and $NT$ {(or, more generally, $MNT$)} in the lower layer. However, the final estimators only involve $S=NT$ samples. {With $M>1$, the final estimators would involve $S=MNT$ samples.} 
{
\subsection{About the choice of the denominator $\bm{\Phi}(\x)$}
}

The computation of the weights in the lower layer allows for different possible denominators, shown in Table \ref{tab:denoLAIS}. 
The function $\bm{\Phi}(\x_{n,t})$ can be taken to be the proposal that actually generated $\x_{n,t}$ ({\it standard}), the mixture of proposals across different chains ({\it spatial}), the mixture of proposals within the chain ({\it temporal}), or the mixture of all proposals ({\it complete}).
Note that, we always have the evaluation of the complete posterior in the numerator, hence all the weighting strategies have the same number of posterior evaluations, i.e., $NT$. However, in practice, 
the cost of the {\it complete}, {\it temporal} and {\it spatial} weighting schemes is higher than the {\it standard} one, and it will increase the overall computation time. This is more obvious in real applications where many chains are run for a long time, i.e., $T$ and $N$ are very large. Commonly, $T\gg N$, so that the {\it spatial} scheme is cheaper than the {\it temporal} scheme, and both are much cheaper than the {\it complete} scheme. In return, these schemes can produce a {remarkable} improvement in the performance of the final estimators. { It can be proved theoretically that the deterministic mixture denominators produce estimators with lower (or equal) variance than the standard weighting \cite{elvira2015generalized}.}
Indeed, our experiments in Section \ref{sec_second_exp} show that the complete denominator consistently produces more stable estimators with only a small increase in computational cost, as compared to the overall cost of the algorithm.
%
%
{
\subsection{Elements for the design of a specific LAIS implementation}
}
 It is important to note that a specific implementation of LAIS is determined by the choices of 
\begin{enumerate}
\item the invariant densities $p_n({\bm \mu})$; 
\item  the MCMC approach (e.g., parallel or single longer chain Metropolis-Hastings, advanced MCMC schemes, etc.), 
\item the proposals $q_{n,t}(\x|\bm{\mu}_{n,t})$; and 
\item  the denominator $\bm{\Phi}(\x)$. 
\end{enumerate}
We define a specific LAIS implementation with a particular choice of those four elements. Below, we present several variants and improvements for the LAIS framework concerning each one of the elements above. For instance, regarding the pdfs  $p_n({\bm \mu})$, we describe the suitable use of different type of  tempered posteriors. The application of sophisticated MCMC algorithms in the upper layer is also discussed.  Recycling sample schemes (which involve the selection of proposals $q_{n,t}$ as well)  and the design of cheap denominators $\bm{\Phi}$ in the lower layer are also introduced in the next sections.





{
\section{Data tempering and partial posteriors as  $p_n({\bm \mu})$} \label{DataTempering}
}


{In the LAIS framework, we have the flexibility in the upper layer design of selecting different invariant densities $p_n({\bm \mu})$. A theoretical discussion regarding the optimal choice of the upper layer densities is given in Appendix \ref{Choiceofpmu}. In this section, we introduce the possibility of using {\it partial posteriors} (i.e., posteriors considering a reduced number of data) as invariant pdfs $p_n({\bm \mu})$. The benefit is twofold: (a) reducing the cost of the posterior evaluations in the upper layer and (b) helping the space exploration of MCMC chains. This second effect is often called {\it data tempering}.\footnote{See Appendix \ref{TemperingSect} for further details.}}
\newline
Specifically, let $\y_n \in\mathbb{R}^{K_n}$ denote a subset of data points, i.e., $\y_n \subset {\bf y}_{\texttt{tot}}$ (with $K_n \ll D_Y$) and assume we have $N$  subsets $\y_1,\dots,\y_N$. For the sake of simplicity, we assume that $\{\y_n\}_{n=1}^N$ represents a partition of ${\bf y}_{\texttt{tot}}$, i.e., $N$ non-overlapping pieces such that $\sum_{n=1}^NK_n=D_Y$. However, more generally, we could also have $\y_n \cap \y_{n'} \neq \emptyset$.\footnote{Note that we are keeping the vector notation for data subset $\y_n$ but sometimes we use it as a set notation, just for the sake of simplicity.  }
We can define the partial posteriors and use them as invariant densities in the upper layer, 
\begin{align}
	p_n({\bf x})=\bar{\pi}_n({\bf x}|{\bf y}_n)\propto L_n({\bf y}_n|{\bf x}) g_n({\bf x}),
\end{align}
where $L_n(\y_n|\x)$ is the likelihood of the batch $\y_n$, and $g_n({\bf x})$ plays the role of a partial prior pdf. For our purpose, we can keep $g_n({\bf x})=g({\bf x})$ for all $n$, or we can split the prior contribution into each data subset, for instance, setting $g_n(\x) = g(\x)^{\frac{1}{N}}$ for all $n$, which is a typical choice in several settings \cite{Scott2013,Neiswanger2014}. 
 Therefore, the partial posterior $\post_n$ is a tempered version of the posterior since its likelihood $L_n(\y_n|\x)$ is less informative, i.e., wider, than in the case where we consider all data.
\newline
Thus, we consider that each MCMC chain in the upper layer addresses a different partial posterior $p_n({\bf x})=\post_n(\x|\y_n)$ ($n=1,\dots, N$). Hence, there are as many chains as number of partial posteriors. We call this scheme  as {\it partial posteriors LAIS} (PLAIS) method.
Note that, in PLAIS, we still evaluate the complete posterior in the lower layer, so the total number of full posterior evaluations is $NT$ (in the lower layer).
\newline
Furthermore, the use of partial posteriors produces more dispersed location parameters of the proposals in the lower layer.  This increases the robustness of the method, since it reduces the chance of obtaining huge weight values and, as a consequence, avoids IS estimators with infinite variance (see the example 1 in \cite{llorente2020marginal}).

%


{
\section{Hamiltonian and Gibbs-driven importance samplers}\label{UppAS}
}


The simplest choice of MCMC schemes in the upper layer is a unique Metropolis-Hastings (MH) chain, or  to employ $N$ independent parallel MH algorithms. However, more sophisticated algorithms can be considered (such as Langevin, Hamiltonian and Gibbs samplers), which can further enhance the performance of the algorithm. On the other hand, the LAIS approach can be interpreted as a way to help these MCMC schemes to improve their efficiency and allow them to estimate efficiently the marginal likelihood $Z$ {(as shown in the numerical experiments in Section \ref{sec_experiments}).}
\newline
\newline
{\bf Hamiltonian MC in the upper layer.}
The Hamiltonian Monte Carlo (HMC) algorithm is usually considered as the state-of-the-art technique in the MCMC world. However, as with the rest of MCMC methods,  it is not straightforward  to estimate the marginal likelihood with HMC samples \cite{llorente2020marginal}. Additionally, it is well-known the difficulty of tuning its hyperparameters for obtaining  efficient sampling \cite{livingstone2019robustness}. In this context, we propose using different HMC algorithms in the upper layer in Table \ref{tab:LAIS_algorithm} , each chain employing possibly  different parameters. Thus, several sets of parameters are jointly used. Note also that we do not need to fine-tune the hyperparameters since the states in the upper layer are not used directly as samples in our framework.  The lower layer in LAIS provides a straightforward estimation of the marginal likelihood.
We compare the performance of these algorithms, denoted as HMC-LAIS, with HMC in Sect. \ref{sec_second_exp}.
\newline
\newline
{\bf Gibbs algorithms in the upper layer.}
Another possibility is to use Gibbs samplers in the upper layer \cite{Robert04}. 
Considering the use of full-conditionals, the Gibbs sampler can be slow since it is a  component-wise scheme, i.e., each component of the parameter vector $\x$ is drawn from the corresponding full-conditional keeping fixed the rest of components.    
However, they have the advantage of working in lower dimension, which allows for designing more efficient samplers in high dimensional spaces.
For instance, extremely efficient MH-within-Gibbs algorithms can be designed using Adaptive Rejection Metropolis schemes for drawing from each one-dimensional full-conditional  \cite{Gilks95,Meyer08,MARTINOfuss}. {This is particularly useful for drawing from very tight posteriors, as shown in \cite{MARTINOfuss} (see also Section \ref{ChaoticSystem}).} 
Other possibility is to employ the adaptive direction sampling which can speed up the mixing of Gibbs chains, choosing different one-dimensional direction of sampling at each iteration \cite{gilks1994adaptive}.
\newline
\newline
More generally, the joint use of HMC, Langevin and Gibbs-based schemes can be potential applied in the upper layer. Note that HMC-LAIS and Gibbs-LAIS are very useful schemes for sampling from concentrated/tight posteriors or high-dimensional posteriors.  
{ 
\subsection{Optimizers versus samplers}
 Let us consider for simplicity the choice $p_n({\bm \mu})=\bar{\pi}({\bm \mu})$ suggested in \cite{martino2016layered}. A simpler alternative could be simply to perform optimization steps for obtaining the location parameters ${\bm \mu}_i$. 
 However, a sampler takes into account not just the modes of $\bar{\pi}({\bm \mu})$ but all the probability mass around these modes. Therefore, using a sampler,  location parameters  $\bar{\pi}({\bm \mu})$ would be spread in the regions of high probability mass (not only at the modes; or close to the modes). This aspect ensures and induces robustness in the IS scheme which uses proposal densities with location parameters ${\bm \mu}_i$, since the full-mixture of proposal densities tends to have a greater variances than the variance of posterior distribution. See Appendix \ref{Choiceofpmu}, for further details. This property is extremely important since avoids the catastrophic scenario of  infinite variance in the final IS estimators, which can occurs  when the proposal density has smaller variance than the target pdf (see the illustrative example 1 in \cite{llorente2020marginal}).  
}


{
\subsection{Upper layer design: a summary}

So far (in Sections \ref{DataTempering} and \ref{UppAS}), we have proposed strategies for improving the efficiency of the final estimators of LAIS. These enhancements are particularly relevant in different challenging inference scenarios, such as tight posteriors and/or high dimensional problems. 
For other complex settings, such as multimodal posteriors, the use of parallel MCMC chains (already suggested in \cite{martino2016layered}) is important. Table \ref{tab:Scenarios} outlines the correspondence between inference scenarios (as well as other features and benefits) and the proposed procedures to employ in the upper layer. For instance, the data tempering is useful in multimodal and high-dimensional scenarios, and particularly useful in the case of concentrated posteriors. Moreover, the data tempering generally increases the robustness of LAIS. Last but not least, observe that all the techniques can be employed jointly in the upper layer, for instance, parallel HMC (or Gibbs) chains (with different parameters) addressing different partial posteriors. In this sense, LAIS can ensure good and robust performance. See Section \ref{sec_experiments} for further details. 
}

\begin{table}[!h]
{
	\centering
	\caption{\small Table of correspondence between benefits and inference scenarios, versus the  proposed procedures (and methods) in the upper layer ($\checkmark =$ useful, and $\bigstar =$  very useful). }
	\vspace{0.2cm}
	{\small
	\begin{tabular}{|c||c|c|c|c|}
		\hline 
		{\bf Methods/}&{\bf Multimodality/} & {\bf Robustness}  & {\bf concentrated/tight}  & {\bf high}  \\
	{\bf Procedures}	&{\bf  helping} &  {\footnotesize{\bf(e.g., to the choice}}  & {\bf posteriors}    &{\bf dimensional} \\
{\footnotesize{\bf (upper layer)}}		&{\bf the exploration}  &  {\footnotesize{\bf of proposal parameters)}}  &   & {\bf spaces}  \\
		\hline
parallel chains	& $\bigstar$ & $\bigstar$  & & \\ 
data-tempering	& $\checkmark$ &  $\checkmark$ & $\bigstar$ &  $\checkmark$ \\
HMC-driven && & $\checkmark$ & $\bigstar$  \\
Gibbs-driven && &  $\bigstar$ & $\checkmark$ \\
		\hline
	\end{tabular}
	}
	\label{tab:Scenarios}
}
\end{table}

{
\section{Compression for parsimonious sampling and weighting}\label{sec_compression}
}

The {\it complete} weighting scheme (see Table \ref{tab:denoLAIS}) provides the best performance in terms of variance, at the expense of an increase in the computational cost, specially in real applications since $T$ and $N$ can be very large. One possibility in order to reduce this cost, without decreasing $T$ or $N$, is the use of partial MIS denominators \cite{elvira2015generalized}. 
 Another approach consists in using some technique that summarizes the population of $NT$ samples. A first attempt has been provided in \cite{el2019efficient}. Another possible way is to apply  a compression  of  Monte Carlo samples  \cite{martino2021compressed}, as we describe below.
These schemes reduces the cost of both sampling and weighting in the lower layer.
\newline
{\bf Compressed LAIS (CLAIS).} Let consider a set of $R$ means $\left\{{\bm \mu}_{k}\right\}_{k=1}^R$ generated by MCMC in the upper layer, and let $B$ be a constant value such that $B<R$. Note that, in the case of $N$ parallel chains of length $T$ in the upper layer, we have $R=NT$.
Given a partition of $\mathcal{X}_{\texttt{tot}}$, i.e., $\mathcal{X}_{1} \cup \mathcal{X}_{2} \cup \ldots \cup \mathcal{X}_{B}=\mathcal{X}_{\texttt{tot}}$ formed by convex, disjoint sub-regions $\mathcal{X}_{m}$,\footnote{{The partition $\mathcal{X}_{1} \cup \mathcal{X}_{2} \cup \ldots \cup \mathcal{X}_{B}=\mathcal{X}_{\texttt{tot}}$ can be obtained using some a-priori information or, as an example, by means of a clustering method.}} we denote the subset of the set of indices $\{1, \ldots, R\}$,
$$
\mathcal{J}_{m}=\left\{i=1, \ldots, R: {\bm \mu}_{i} \in \mathcal{X}_{m}\right\}, \qquad { m=1,...,B,}
$$
which are associated with the samples in the $m$-th sub-region $\mathcal{X}_{m}$. 
The cardinality $\left|\mathcal{J}_{m}\right|$ denotes the number of samples in $\mathcal{X}_{m}$ and we have $\sum_{m=1}^{B}\left|\mathcal{J}_{m}\right|=R$.
 We can compress the information contained in samples, constructing a stratified approximation based on $B$ weighted particles $\left\{\mathbf{s}_{m}, {a}_{m}\right\}_{m-1}^{B}$, 
where $\mathbf{s}_{m}$ is a (properly chosen) point in $\mathcal{X}_m$ and ${a}_{m}=\frac{\left|J_{m}\right|}{R}$. 
\newline
\newline
{\bf Possible choices of $\mathbf{s}_{m}$.} The summary points  $\mathbf{s}_{m}$ can be randomly chosen, picking uniformly a mean in $\mathcal{X}_m$,  in the set $\{{\bm \mu}_{i} \}_{i\in \mathcal{J}_{m}}$ or using a deterministic procedure,  e.g.,
\begin{equation}\label{s_deter}
{\bf s}_{m}=\frac{1}{|J_{m}|}\sum_{j \in J_{m}}  {\bm \mu}_{j}. 
\end{equation}
For the statistical properties of these choices  see \cite{martino2021compressed}. Other choices based on empirical quantiles are also possible.
As an example, a suitable compression scheme can be provided applying a clustering method to the set  $\left\{{\bm \mu}_{k}\right\}_{k=1}^R$, where $M$ represents the number of clusters.  
After the compression, we can consider as proposal and denominator in the lower layer the following mixture of densities $p(\x|{\bf s}, {\bm \Sigma})$ where ${\bf s}$, $ {\bm \Sigma}$ represent a location parameter  and a covariance matrix,
\begin{equation}
q_B({\bf x})=\sum_{m=1}^{B} {a}_{m} p\left(\mathbf{x}| \mathbf{s}_{m}, {\bm \Sigma}\right).
\end{equation}
 Thus, the mixture $q_B$ is used for sampling and computing the weights in the lower layer.
 A suitable choice of $ \mathbf{s}_{m}$ and ${\bm \Sigma}$ is the key point for the success of the compressed scheme. For the summary points  $\mathbf{s}_{m}$, we suggest the deterministic procedure in Eq. \eqref{s_deter}. 
 \newline
 \newline
{\bf Suitable choice  of ${\bm \Sigma}$.}  We suggest to obtain the $D_X \times D_X$ covariance matrix ${\bm \Sigma}$ as 
\begin{eqnarray}\label{eq_Sigma}
\boldsymbol{\Sigma}&=&{\bf Q}_{\mu}-{\bf Q}_C+\sigma_p^2 \mathbf{I}.
\end{eqnarray}
where ${\bf Q}_{\mu}=\frac{1}{R}\sum_{k=1}^R \left({\bm \mu}_{k}-{\bf m}\right)\left({\bm \mu}_{k}-{\bf m}\right)^{\top}$ with ${\bf m}=\frac{1}{R}\sum_{k=1}^R{\bm \mu}_{k}$ is the covariance matrix of all $R$ means ${\bm \mu}_k$,
 and ${\bf Q}_C=\sum_{m=1}^M a_m\left({\bf s}_{m}-{\bf m}_C\right)\left( {\bf s}_{m}-{\bf m}_C\right)^{\top}$ with ${\bf m}_C=\sum_{m=1}^M a_m {\bf s}_{m}$ is the covariance matrix of the summary samples.\footnote{Clearly, if ${\bf s}_m$ are chosen as in Eq. \eqref{s_deter}, then ${\bf m}={\bf m}_C$.} Finally, $\sigma_p^2$ is chosen by the user. 
   With  ${\bf s}_m$ in Eq. \eqref{s_deter}, it is possible to show that 
 \begin{equation}
 {\bf Q}_{\mu}-{\bf Q}_C= \sum_{m=1}^B a_m \left(\frac{1}{|J_{m}|}\sum_{j \in \mathcal{J}_{m}} \left({\bm \mu}_{j}-\mathbf{s}_{m}\right)\left({\bm \mu}_{j}-\mathbf{s}_{m}\right)^{\top}\right).
\end{equation}
That is, the covariance of each component in $q_M$ is the weighted average of the covariances within clusters plus the term $\sigma_p^2{\bf I}$. In the following, we explain the reason of using Eq. \eqref{eq_Sigma}. { We remark that a suitable choice of $\boldsymbol{\Sigma}$ is crucial for the performance of the compression technique. The proposed covariance matrix $\boldsymbol{\Sigma}$ in Eq. \eqref{eq_Sigma} is a robust choice which provides good performance, as shown in Section \ref{sec_second_exp}. } 
\newline
The combined choice of ${\bf s}_m$ in Eq. \eqref{s_deter}  and ${\bm \Sigma}$  in \eqref{eq_Sigma} has the following property.
Let us assume that, without compression, we would like to use $B$ proposal densities $q$ in the lower layer with a covariance matrix $\sigma_p^2{\bf I}$.
  Without compression, we have $B=R$, ${\bf s}_k={\bm \mu}_k$, ${\bf Q}_{\mu}={\bf Q}_C$, so we have the covariance of each mixture component is ${\bm \Sigma}=\sigma_p^2{\bf I}$, as expected. 
   With the maximum compression, $B=1$, then ${\bf Q}_C$ is null and  ${\bm \Sigma}={\bf Q}_{\mu}+ \sigma_p^2{\bf I}$. Hence, with maximum compression, the proposal $q_B$ takes into account the dispersion set by the user (by the term $\sigma_p^2 {\bf I}$) plus the covariance matrix of the $R$ means ${\bm \mu}_k$ (i.e., the term ${\bf Q}_{\mu}$), obtained in the upper layer.
  Note that, clearly, the cost of the  employed  compression technique must be lower than the cost of evaluating the full denominator.  We test the performance of CLAIS with several choices of $R$, and compare it with standard LAIS  in Section \ref{sec_second_exp}.


\section{Recycling LAIS (RLAIS)}\label{RLAISsect} 

In this Section, we discuss the possibility of recycling the samples, and their corresponding evaluations, from the upper layer for their use in the lower layer, hence reducing the overall computational cost.
For simplicity, let us assume the use of $N$ parallel Metropolis-Hastings (MH) algorithms in the upper layer.
Moreover, in this first part of the section,  assume that $p_n=\post$ for all $n$. 
Given the initial state $\bm{\mu}_{n,0}$, a proposal pdf $\varphi_n$, and a length value $T$, 
the $n$-th MH chain follows the following steps:
\newline
		\begin{tabular}{p{0.95\columnwidth}}
		\normalsize
		\\ 
		\normalsize
		- {\bf For $t=1,\ldots,T$:}
		\begin{enumerate}
			\normalsize
			\item  Draw ${\bf z}_{n,t}\sim \varphi_n({\bf x}|\bm{\mu}_{n,t-1})$.
			\item  Set $\bm{\mu}_{n,t}={\bf z}_{n,t}$ with probability
			\begin{equation}\label{eq:MHaccprob}
				\alpha=\min\left[1,\frac{\pi({\bf z}_{n,t}|{\bf y}_{\texttt{tot}})\varphi_n(\bm{\mu}_{n,t-1}|{\bf z}_{n,t})}{\pi(\bm{\mu}_{n,t-1}|{\bf y}_{\texttt{tot}})\varphi_n({\bf z}_{n,t}|\bm{\mu}_{n,t-1})}\right],
			\end{equation}
			otherwise, set $\bm{\mu}_{n,t}=\bm{\mu}_{n,t-1}$ (with probability $1-\alpha$).
		\end{enumerate}
		-  {\bf  Outputs:} The chain $\{\bm{\mu}_{n,t}\}_{t=0}^{T-1}$. Additionally, we obtain and store $\{{\bf z}_{n,t}\}_{t=1}^T$, $\{\pi({\bf z}_{n,t}|{\bf y}_{\texttt{tot}})\}_{t=1}^T$ and $\{\varphi_n({\bf z}_{n,t}|\bm{\mu}_{n,t-1})\}_{t=1}^T$.
		\\  \\
	\end{tabular}
	
\noindent
Therefore, at each iteration, a candidate is drawn ${\bf z}_{n,t} \sim \varphi_n(\x|\bm{\mu}_{n,t-1})$ and then it is tested (accepted or discarded) as possible new state, according to the  acceptance MH probability. 
If we store all candidates $\{{\bf z}_{n,t}\}_{t=1}^T$  and the corresponding evaluations of the posterior  $\{\pi({\bf z}_{n,t}|{\bf y}_{\texttt{tot}})\}_{t=1}^T$ (for all $n$), required in the computation of $\alpha$ in Eq. \eqref{eq:MHaccprob}, we can use them in the lower layer as samples, i.e., we set $\x_{n,t-1} = {\bf z}_{n,t}$. In this way, we reduce the computation time since we do not need to draw additional samples.  
\newline
Note that $\varphi_n(\x|\bm{\mu}_{n,t-1})$ becomes the proposal in the lower layer, i.e., we set $q_{n,t}(\x)=\varphi_{n}(\x|\bm{\mu}_{n,t-1})$. 
The evaluations of the proposal $\varphi_n({\bf z}_{n,t}|\bm{\mu}_{n,t-1})$ can be also stored.  Depending on the choice of the weighting scheme, other evaluations of different proposals $\varphi_j$, with $j\neq n$ can be required. { This also produces  a slight reduction of the cost of evaluating the denominator of the weights in the lower layer. See the next section for further details.}
The algorithm is outlined in Table \ref{tab:LAIS_with_subpost_Recycling}, and Table \ref{tab:deno2} shows different weighting procedures. 
Since  $p_n=\post$  and the posterior evaluations  are recycled, the total number of posterior evaluations in RLAIS is only $E=NT$.
\newline
{\bf Consistency.}  It is important to note that we can find an equivalent proposal $\widetilde{q}_{MH}({\bf x})$ of MH-type algorithms which can be expressed as a convolution integral, similarly as we have done in LAIS. See the Appendix \ref{HI_MH} for more details. 
In RLAIS, the different MIS denominators can be considered as Monte Carlo approximations of this equivalent proposal $\widetilde{q}_{MH}$, expressed as an integral in Eq. \eqref{superEq}. Therefore, in the case of the first $3$ different MIS denominators (the complete, spatial and temporal mixtures) as $N$ and $T$ grow, the chosen denominator provides an better approximation of the $\widetilde{q}_{MH}$ and the MIS weights becomes closer and closer to standard importance weights of the form $w_{n,t}=\frac{\pi({\bf x}_{n,t}|{\bf y}_{\texttt{tot}})}{\widetilde{q}_{MH}({\bf x}_{n,t})}$.  
RLAIS can be seen as a multiple-chain generalization of \cite{schuster2020markov,rudolf2020metropolis}.



\begin{table}[!h]
	\centering
	\caption{LAIS with recycling of samples (RLAIS)}
	\begin{tabular}{|p{0.95\columnwidth}|}
		\hline
		\begin{itemize}
			\item[1.] {\bf Sampling:} 
			 Let consider Metropolis-Hastings (MH)-type schemes with random walk proposal densities $\varphi_{n,t}({\bf x}|{\bm \mu}_{n,t})$ ($\varphi_{n,t}$ can vary with $t$ since we assume they can be also adaptive schemes),  generating $N$ MCMC chains of length $T$.
		
Then, the states of the chains are  ${\bm \mu}_{n,t}$,  for $n=1,\ldots,N$ and $t=1,...,T$. At each iteration of  one MH scheme, we draw a candidate ${\bf z}_{n,t}\sim\varphi_{n,t}({\bf x}|{\bm \mu}_{n,t-1})$ that will be accepted or rejected in the MH step. We save all the $NT$ candidates ${\bf z}_{n,t}$ for $n=1,\ldots,N$ and $t=1,...,T$.
			
			\item[2.] {\bf Weighting:}   Assign to ${\bf z}_{n,t}$ the weights
			\begin{equation}
				w_{n,t}=\frac{\pi({\bf z}_{n,t}|{\bf y}_{\texttt{tot}})}{{\bm \Psi}({\bf z}_{n,t})},
			\end{equation}
			where  different possible choices for ${\bm \Psi}({\bf z}_{n,t})$ are possible (see Table \ref{tab:deno2}). 
			\item[3.] {\bf Output:} Return all the pairs  $\{{\bf z}_{n,t},w_{n,t}\}$,  and/or the estimators given in Eqs \eqref{eq:Z_EST} and \eqref{eq:I_EST}.
		\end{itemize}\\
		\hline
	\end{tabular}
	\label{tab:LAIS_with_subpost_Recycling}
\end{table}

\begin{table}[!h]
	\centering
	\caption{Possible denominators ${\bm \Psi}({\bf x}_{n,t})$.}
	\begin{tabular}{|c|c|c|c|}
		\hline 
		{\bf complete} & {\bf temporal}  & {\bf spatial}  & {\bf standard}  \\
		\hline
		& & & \\
		$\frac{1}{NT}\sum_{\tau=0}^{T-1}\sum_{n=1}^N \varphi_{n,\tau}({\bf x}_{n,t}|{\bm \mu}_{n,\tau})$    & $\frac{1}{T}\sum_{\tau=0}^{T-1}\varphi_{n,\tau}({\bf x}_{n,t}|{\bm \mu}_{n,\tau})$ & $\frac{1}{N}\sum_{n=1}^N \varphi_{n,t}({\bf x}_{n,t}|{\bm \mu}_{n,t})$ &  $\varphi_{n,t}({\bf x}_{n,t}|{\bm \mu}_{n,t})$   \\
		& & &  \\
		\hline
	\end{tabular}
	\label{tab:deno2}
\end{table}

\noindent
{\bf PLAIS with recycling (PA-RLAIS).} We can combine the idea of using the partial posteriors and the RLAIS approach.
Indeed, also in PLAIS, it is possible to avoid the sampling step if we recycle all candidates produced within the MH algorithms in the upper layer. We denote the resulting scheme as PA-RLAIS.
 We can recycle the candidates  $\{{\bf z}_{n,t}\}_{t=1}^T$  and the proposal evaluations $\{\varphi_n({\bf z}_{n,t}|\bm{\mu}_{n,t-1})\}_{t=1}^T$ (for all $n$) but, in this scenario, we have not evaluations of the full posterior in the upper layer (then we cannot recycle the posterior evaluations).

{
\section{Computation costs of the proposed schemes}\label{sec_compt_costs}
}

%
%
%
Generally, the most costly step is the evaluation of the complete posterior $\pi(\x|{\bf y}_{\texttt{tot}})$ (due to a costly model or number of data). The evaluation of the partial posteriors is not that costly since we choose the batch sizes such that $K_n\ll D_Y$ for all $n=1,\dots, N$. Thus, the comparison among PLAIS, RLAIS and PAPIS, as well as with other methods, must be done in terms of number of evaluations of the (unnormalized) posteriors, the complete posterior $\pi(\x)$, and/or the partial posteriors $\pi_n(\x)$'s. 
 A summary of the number of evaluations of $\pi(\x)$ and all partial posteriors $\pi_n(\x)$'s is given below:

	\begin{center}
	\begin{tabular}{|c|cc|c|c|}
		\hline
		\multirow{2}{*}{{\bf Method}} 
		& \multicolumn{2}{c|}{{\bf Upper layer}} & {\bf Lower layer} & {\bf Drawing samples} \\
		& evals of $\pi(\x|{\bf y}_{\texttt{tot}})$ & evals of $\pi(\x|\textbf{y}_n)$ &  evals of $\pi(\x|{\bf y}_{\texttt{tot}})$  & {\bf in the lower layer}  \\
		\hline
		LAIS    &  $NT$ &  0 &  $NT$ &   \checkmark  \\
		PLAIS  & 0 & $NT$ &    $NT$ &   \checkmark  \\
		RLAIS & $NT$ & 0   &  0 &   \text{\sffamily X}   \\
		PA-RLAIS & 0 & $NT$   &  $NT$  &   \text{\sffamily X} \\
		\hline 
		\hline
		--- & --- &   cheaper &---  & --- \\
		\hline
		\hline
		\multicolumn{5}{|c|}{{\small CLAIS can be also combined with the other schemes above for building cheaper denominators.}} \\
		\hline
	\end{tabular}
\end{center}

\noindent
{   Therefore, the total number of full-posterior evaluations of the standard LAIS scheme is $E=NT+NT=2NT$. If we draw $M>1$ samples from each proposal density $q_{n,t}$ in the lower layer, the total number of full-posterior evaluations would be $E=NT+MNT=(M+1)NT$.
\newline
\newline
If we  denote as $C$ the {\it atomic cost} of evaluating once the likelihood function with only one data, then the total  cost associated to the total number of the target evaluations (considering  evaluations of full-posterior and/or evaluations of partial posteriors) of different techniques is given below:
\begin{center}
	\begin{tabular}{|c|l|}
		\hline
		\multirow{2}{*}{{\bf Method}}  & \multirow{2}{*}{{\bf Total cost associated to the posterior evaluations}}  \\
		 &  \\
		\hline
		\multirow{2}{*}{LAIS}    &   \multirow{2}{*}{$\texttt{Cost}=2NTCD_Y$} \\
		 & \\
		\hline 
		\multirow{2}{*}{PLAIS}  &   $\texttt{Cost}=TC\left(\sum_{n=1}^N K_n\right) + NTCD_Y$   \\
		&  $\texttt{Cost} = TCD_Y + NTCD_Y = (N+1)  TCD_Y $\\
		\hline
		\multirow{2}{*}{RLAIS} &   \multirow{2}{*}{$\texttt{Cost}=NTCD_Y$}  \\
		& \\
		\hline 
		\multirow{2}{*}{PA-RLAIS} &  \multirow{2}{*}{$\texttt{Cost} = TCD_Y + NTCD_Y= (N+1)  TCD_Y$ }\\
		& \\
		\hline 
	\end{tabular}
\end{center}
where $N$ is the number of chains (with length $T$) in the upper layer, $D_Y$ is the total number of data, and $C$ is the atomic cost previously described.  We have used that $\sum_{n=1}^N K_n=D_Y$ where  $K_n$ are the number of data in the $n$-th partial posterior. Clearly, RLAIS and standard LAIS are the algorithms with lowest and greatest costs, respectively, as shown below.
\begin{center}
\begin{tabular}{|l|}
		\hline
		{\bf Inequalities in terms of cost of total posterior evaluations:}  \\
		\hline
		$\texttt{Cost of RLAIS} < \texttt{Cost of PA-RLAIS}=\texttt{Cost of PLAIS} <  \texttt{Cost of LAIS}$ \\
		\hspace{1cm} $\Downarrow$ \hspace{3cm}  $\Downarrow$ \hspace{3.2cm}  $\Downarrow$ \hspace{3cm}  $\Downarrow$ \\
		\hspace{0.3cm} $NTCD_Y$  \hspace{0.7cm} $<$ \hspace{0.2cm}  $(N+1)TCD_Y$ \hspace{0.24cm}  $=$  \hspace{0.05cm} $(N+1)TCD_Y$ \hspace{0.01cm}$<$ \hspace{0.2cm} $2NTCD_Y$  \\
		\hline
	\end{tabular}
\end{center}

\noindent
However, considering also the cost of sampling from the proposal pdfs, PA-RLAIS is less costly than PLAIS since it does not require extra samples in the lower layer. This is an additional advantage of RLAIS as well. 
We  recall that the reason of using partial posteriors is not only a reduction on the computational cost. Indeed, the use of partial posteriors fosters the space exploration by the data-tempering effect.
Finally, we also remark that the overall computational cost also depends on the denominator choice: this is the reason of employing the proposed scheme in Section \ref{sec_compression}, denoted as CLAIS. The number of proposal evaluations per sample {\it in the lower layer} with the different possible denominators is given below:
\begin{center}
\begin{tabular}{|c|c|c|c|c|}
		\hline 
		{\bf Method}  &{\bf complete} & {\bf temporal}  & {\bf spatial}  & {\bf standard}  \\
		\hline
		Stand. LAIS & $NT$   & $T$ & $N$ & $1$    \\
		RLAIS & $NT-1$  & $T-1$ & $N-1$ & 0    \\
		\hline
	\end{tabular}
\end{center}
 Recall that, for simplicity,  through this work we have considered to draw $M=1$ sample from each proposal, in the lower layer. However, all the formulas above just suffer some mild changes for $M>1$. 
 }




\section{Numerical experiments}
\label{sec_experiments}

{
In this Section, we test the performance of  the algorithms described in this work. 
We have considered different challenging scenarios. As an example, we tackle multimodal target densities (in Sections \ref{Robustness} and \ref{HighDimEx}), high-dimensional problems (in Section \ref{HighDimEx}) and extremely sharp/tight posteriors (in Section \ref{ChaoticSystem}). In the last experiment (Section \ref{Example covid}), we also analyze real data  in a regression problem on the daily deaths by COVID in Italy. The correspondence between proposed algorithms and sections is given below:
\begin{center}
\footnotesize
\begin{tabular}{|c|c|c|c|c|c|c|}
		\hline 
		{\bf Method}  &{\bf Section \ref{Robustness}} & {\bf Section \ref{sec_first_exp}} & {\bf Section \ref{sec_second_exp}} & {\bf Section \ref{HighDimEx}} & {\bf Section \ref{ChaoticSystem}} & {\bf Section \ref{Example covid}}  \\
		\hline
		Stand. LAIS & $\checkmark$ & $\checkmark$ &&&&     \\
		PLAIS &  & $\checkmark$  &&&&     \\
		CLAIS &  &  & $\checkmark$ &&&     \\
		RLAIS &  &  &&& $\checkmark$ &     \\
		PA-RLAIS &  & $\checkmark$  &&&&     \\
		HMC-LAIS &  &  & $\checkmark$&  $\checkmark$ &&     \\
		Gibbs-LAIS &  &  &&& $\checkmark$ & $\checkmark$     \\
		Diff. Den. ${\bm \Phi}(\x)$ &  &  & $\checkmark$ &&&     \\
		\hline
	\end{tabular}
\end{center}

}


%


{
\subsection{Comparison with benchmark AIS schemes}\label{Robustness}

In this section, we compare LAIS with the most relevant and benchmark AIS schemes proposed in the literature \cite{Cappe04,Cornuet12,ELVIRA201777,APIS15}. The objective of this section is to highlight the robustness of LAIS scheme with respect to the choice of the parameters of method, comparing with the results of the other AIS techniques.
With this aim, we consider a highly-multimodal bivariate target pdf defined as a mixture of five Gaussians, i.e., 
\begin{equation}
\pi(\x)=\frac{1}{5}\sum_{i=1}^5 \mathcal{N}(\x;{\bm \nu}_i,{\bm \Lambda}_i), \quad \x\in \mathbb{R}^2,
\label{Target1b}
\end{equation}
where $\mathcal{N}(\x;{\bm \nu}_i,{\bm \Lambda}_i)$ denotes a  Gaussian density with mean vector ${\bm \mu}_i$ and covariance matrix ${\bm \Lambda}_i$, ${\bf \nu}_1=[-10, -10]^{\top}$, ${\bm \nu}_2=[0, 16]^{\top}$, ${\bm \nu}_3=[13, 8]^{\top}$, ${\bm \nu}_4=[-9, 7]^{\top}$, ${\bm \nu}_5=[14, -14]^{\top}$, ${\bm \Lambda}_1=[2, \ 0.6; 0.6, \ 1]$, ${\bm \Lambda}_2=[2, \ -0.4; -0.4, \ 2]$, ${\bm \Lambda}_3=[2, \ 0.8; 0.8, \ 2]$, ${\bm \Lambda}_4=[3, \ 0; 0, \ 0.5]$, and finally ${\bm \Lambda}_5=[2, \ -0.1; -0.1, \ 2]$. This is a very challenging scenario since we have 5 different modes, far away one from another.  
In this example, we can analytically compute different moments of the target in \eqref{Target1b}, and therefore we can easily validate the performance of the different techniques.
In particular, we consider the computation of the mean of the target, $E[{\bf X}]=[1.6, 1.4]^{\top}$, and the normalizing constant, $Z=1$, for ${\bf X}\sim \frac{1}{Z}\pi(\x)$.
We compute the Mean squared error (MSE) in the estimation of $E[{\bf X}]$ and in the normalizing constant $Z$ (which usually represents a marginal likelihood, when the density of interest is a Bayesian posterior).
\newline
\newline
We apply LAIS with $N$ parallel MH chains in the upper layer (of length $T$). 
We assume Gaussian proposal densities for all of the methods compared, and deliberately choose a bad initialization of the means in order to test the robustness and the adaptation capabilities.
Specifically, the initial location parameters of the  proposals are selected uniformly within the $[-4,4]\times[-4,4]$ square, i.e., ${\bm \mu}_i^{(1)}\sim \mathcal{U}([-4,4]\times[-4,4])$ for $i=1,\ldots,N$. Note that none of the modes of the target are contained within this initialization square.
We test all the alternatives using the same isotropic covariance matrices for all the Gaussian proposals, ${\bf C}_i = \sigma^2{\bf I}_2$, where in some simulations we vary $\sigma$.
All the results have been averaged over $10^3$ independent runs, where the total number of target evaluations $E$ is the same in all the techniques (see Section \ref{sec_compt_costs} for LAIS).  In order to make possible a fair comparison with other schemes, in LAIS we draw $M>1$ samples from each proposal density $q_{n,t}$ in the lower layer, so that  the total number of full-posterior evaluations in LAIS is $E=NT+MNT=(M+1)NT$ (as shown in the previous section).
We apply also the following schemes:  the standar {\bf Population Monte Carlo (PMC)} technique \cite{Cappe04},  the {\bf Adaptive Population Importance Sampling (APIS)} method \cite{APIS15}, 
 the improved PMC schemes {\bf GR-PMC} and {\bf LR-PMC} \cite{ELVIRA201777}, and the {\bf Adaptive Multiple Importance Sampling (AMIS)} approach \cite{Cornuet12}.
We remark that all the comparisons have been performed  with the same number of target evaluations $E$. For instance, in Figure \ref{fig_Ex1_SigN}(a), we vary the standard deviation of the proposal densities $\sigma$, and
we set  $N=10$, $M=9$, $T=100$ for LAIS,  $N=10$, $M=10$ $T=100$ for APIS, GR-PMC and LR-PMC, and $M=100$ and $T=100$ in AMIS (since in AMIS we have a unique proposal density), keeping $E=10^4$. We repeat the experiment in in Figure \ref{fig_Ex1_SigN}(b), but considering $N=100$. In Figure \ref{fig_Ex1_SigN}(c), we set $\sigma=5$ and vary $N$. We can observe that stand. LAIS generally outperforms the other techniques.  Even when LAIS does not provide the smallest MSE, it obtains close results. Namely, LAIS provides competitive results for any of the values $\sigma$ or $N$, proving its robustness.
As $N$ grows, LAIS becomes even more competitive. 

\begin{figure}[!htb]
	\centering 
	\centerline{
		\subfigure[$N=10$]{\includegraphics[scale=0.45]{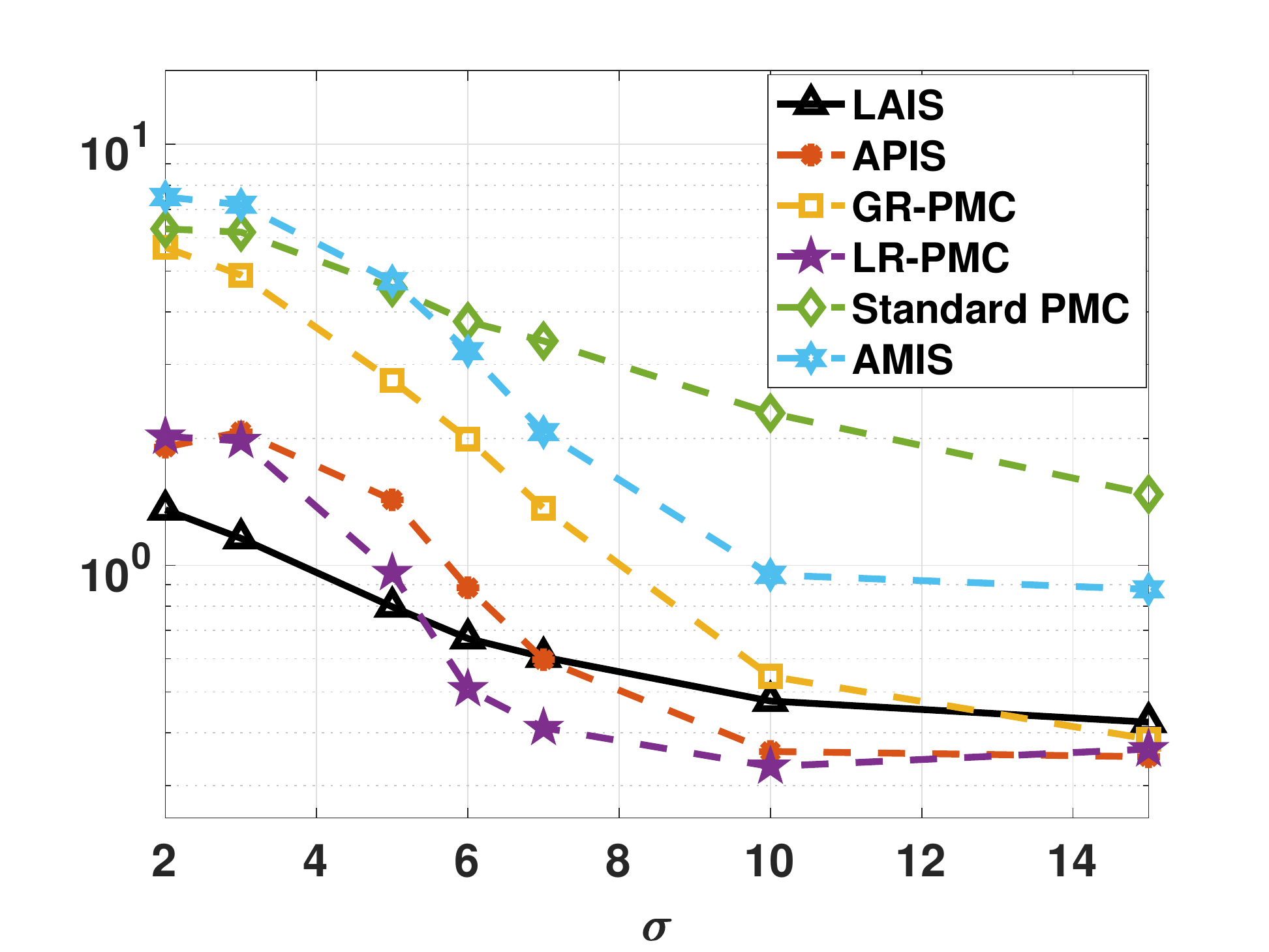}}
		\subfigure[$N=100$]{\includegraphics[scale=0.45]{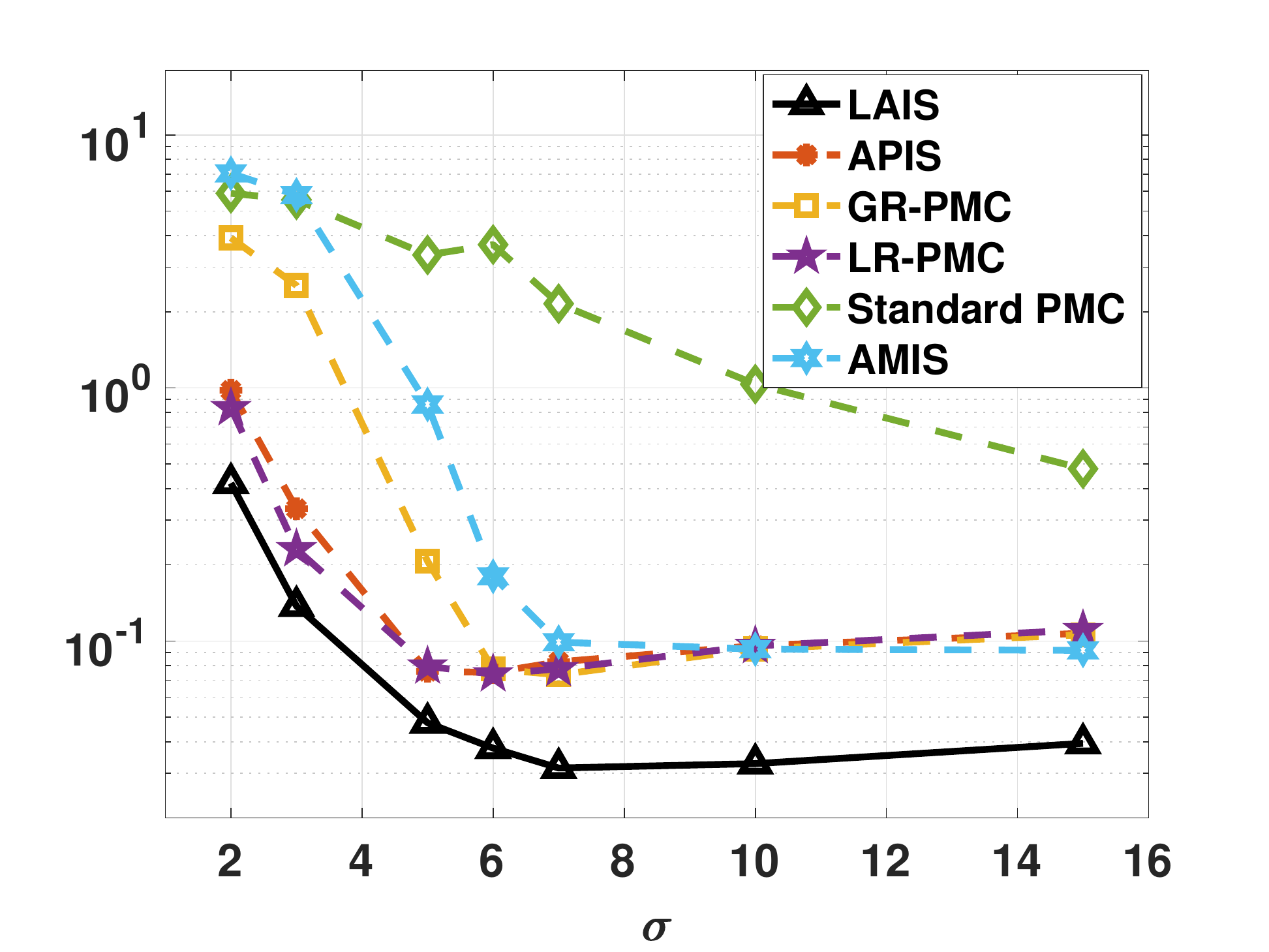}}
	}
	\centerline{
		\subfigure[$\sigma=5$]{\includegraphics[scale=0.45]{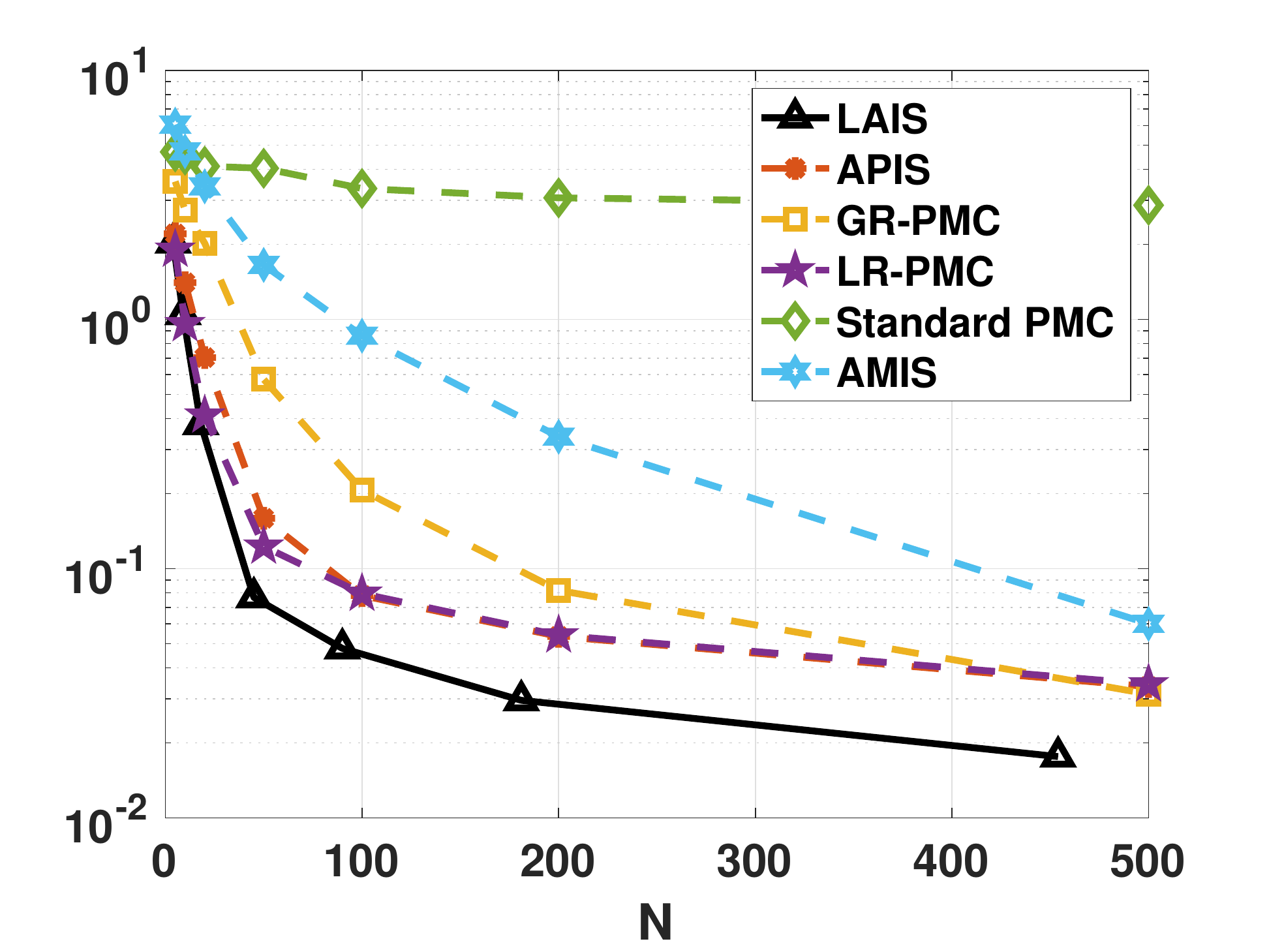}}

	}
	\caption{{ {\bf(Fig. of Section \ref{Robustness})} MSE in log-scale obtained by different techniques, keeping fixed $E=10^4$; {\bf (a)} with $N=10$, and varying $\sigma$; {\bf (b)} with $N=100$ and varying $\sigma$; {\bf (c)} with $\sigma=5$ and varying $N$.   }}
	\label{fig_Ex1_SigN}
\end{figure}

} 
\subsection{Parameter fitting in a non-linear regression problem}
\label{sec_first_exp}
\begin{figure}[!htb]
	\centering 
	\centerline{
		\subfigure[]{\includegraphics[scale=0.4]{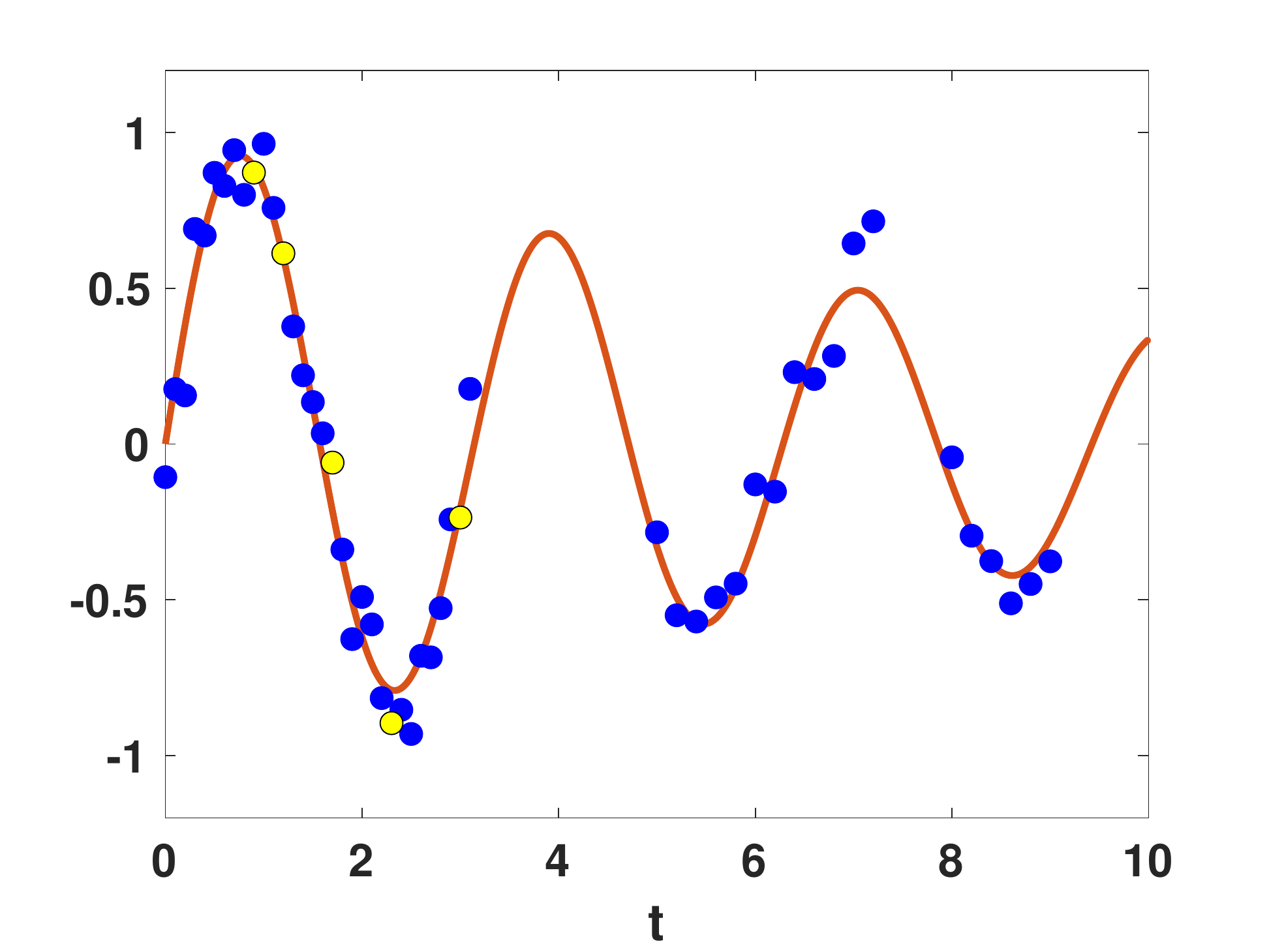}}
		\subfigure[]{\includegraphics[scale=0.4]{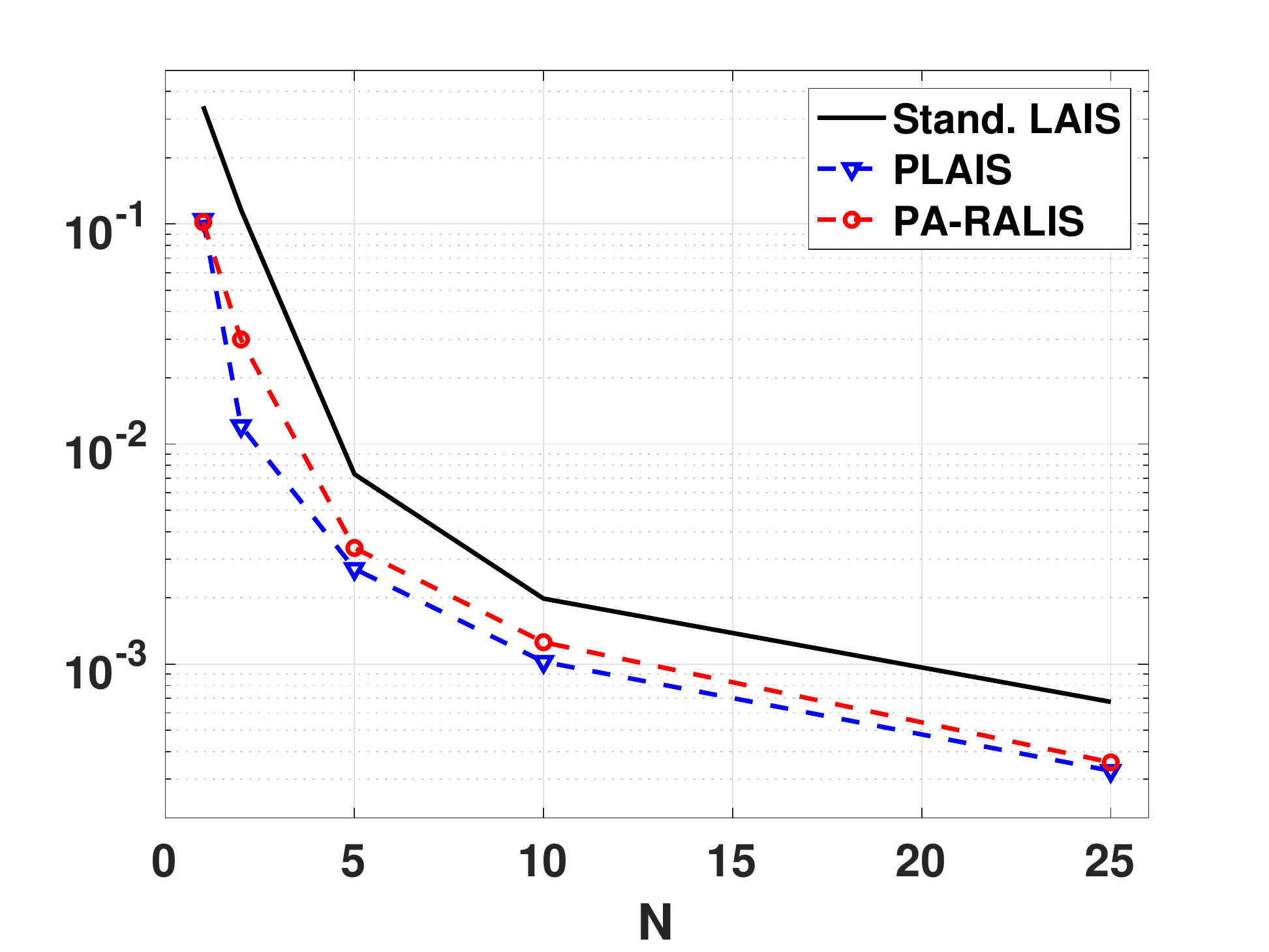}}
	}
	\caption{{{\bf(Fig. of Section \ref{sec_first_exp})}} {\bf (a)} The solid line is the function that defines the model, and the blue dots are the observations generated from it. {The yellow dots represent an example of random subset of data used in a partial-posterior.}{{\bf (b)} MSE versus $N$, with $T=20$ and $K_n=10$ for all $n=1,...,N$.}}
	\label{fig:model and data}
\end{figure}

In this section, we consider a non-linear regression problem. We generate 50 observations, ${\bf y}_\texttt{tot}=\{y_i\}_{i=1}^{50}$, from the following observation model
\[ y_i =  \exp(-\alpha t_i)\sin(\beta t_i) + v_i \]
where the values $ \alpha $ and $ \beta $ were fixed at 0.1 and 2, respectively. 
The error terms $ v_i $ were independently generated from a Gaussian, $ \mathcal{N}(0,0.1^2) $.  
For this model, we take $ \x = [\alpha,\beta]^\top $ and set a uniform density over the rectangle $ [0,10]\times[0,2\pi] $ as prior density for $ \x $.  Figure \ref{fig:model and data}(a) shows the function $\exp(-\alpha t)\sin(\beta t)$ and some data generated according to the model. The goal is to investigate the use of partial posteriors in the LAIS framework when computing  $ \mathbb{E}[\x|{\bf y}_\texttt{tot}] $, $ \mbox{var}[\x|{\bf y}_\texttt{tot}] $ (marginal variances)
and $ Z = p({\bf y}_\texttt{tot}) $.  
By using a very thin grid over the space, we are able to calculate the true values, obtaining $\mathbb{E}[\x|{\bf y}_\texttt{tot}]=[0.1,2]^\top$, $ \mbox{var}[\x|{\bf y}_\texttt{tot}] = [6.88  \cdot 10^{-5} , 8.38\cdot 10^{-5}]^\top $ and $ Z = 3.03\cdot 10^{-15} $.
We compute the MSE in estimating those quantities with the following methods: (a) LAIS, (b) PLAIS, and (c) PA-RLAIS.

 For all the methods, the upper layer consists of $N$ independent random walk Metropolis-Hastings (MH) algorithms with Gaussian proposals (the same for all the schemes). In the upper layer, PLAIS and PA-RLAIS differ from LAIS in that, instead of the full posterior, each of the $N$ chains targets a different partial posterior (with the same number of data $K_n$ for all $n$).
In the lower layer, one sample was drawn from each of the Gaussian proposal pdf.
The covariance matrix of all the Gaussian proposals was set to $ 2{\bf I}_2 $ where ${\bf I}_2$ is a $2\times 2$ unit matrix. 
In the lower layer, PA-RLAIS differs from LAIS and PLAIS, in that no sample is drawn in this second stage, but all samples are recycled from the chains in the upper layer.
 { In a first experiment, we test the values $N\in \{1,2,5,10,25\}$, and set $T=20$, $K_n=10$ for all $n$.  The results (averaged over $10^3$ runs) in terms of MSE are shown in Figure \ref{fig:model and data}(b). We can already see the benefits of PLAIS and PA-RLAIS.  }
 \newline
 {
 In a second experiment, we fix the number of total evaluations of the full-posterior is $E=2000$. In this case, for any value of $N\in \{1,2,5,10,25,50\}$  we change $T$, in order to keep constant the total number evaluations of the full-posterior is $E=2000$ (see Section \ref{sec_compt_costs}). 
 }
In each simulation the partial posteriors were created by choosing randomly $ K_n $ data, with  $ K_n \in \{5,10\} $. 
Figure \ref{fig:model and data}(a) depicts some data generated according to the model. The orange dots are the observations chosen to construct the partial posterior in one simulation with $K_n=5$. 
Finally, in all the methods, the initial mean vectors were drawn from the prior, i.e., $\bm{\mu}_{n,0}\sim \mathcal{U}\left([0,10]\times[0,2\pi]\right)$, for all $n$. The results are averaged over 500 independent simulations.

\begin{figure}[!htb]
	\centering 
	\centerline{
		\subfigure[]{\includegraphics[width=7.5cm]{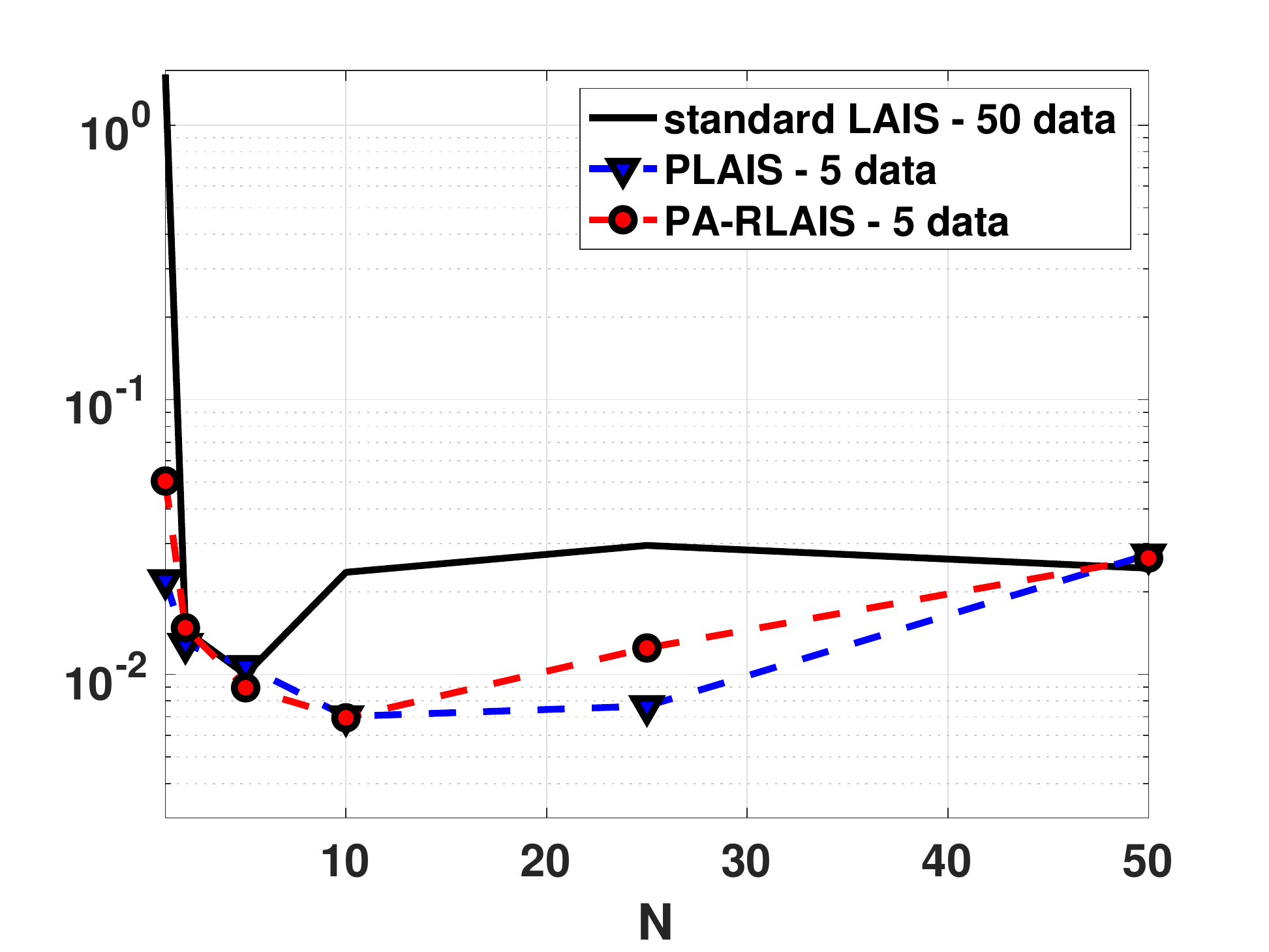}}
		\hspace{0.7cm}
		\subfigure[]{\includegraphics[width=7.5cm]{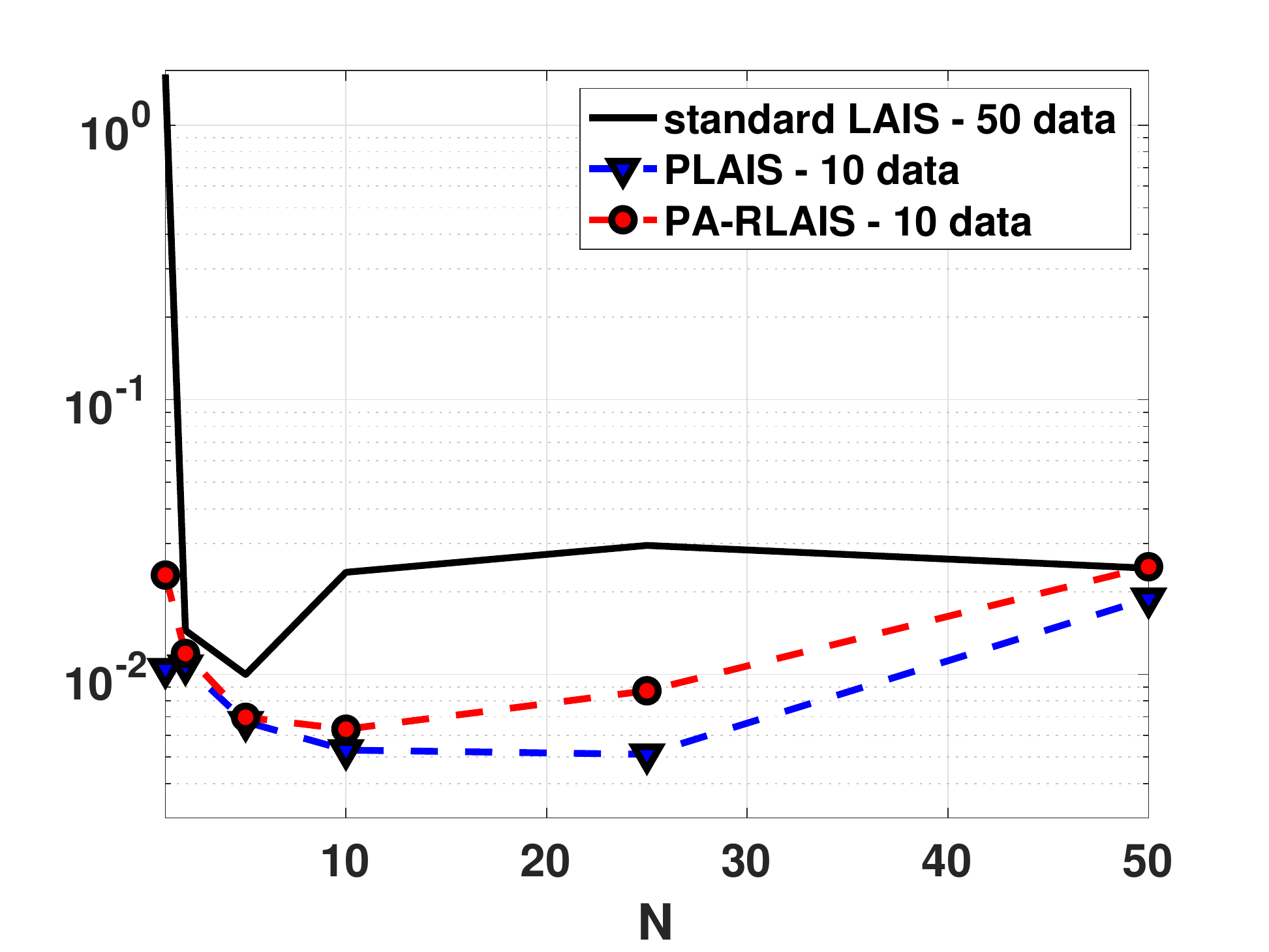}}
	}
	\caption{{{\bf(Fig. of Section \ref{sec_first_exp})}} MSE obtained by the different algorithms for distinct numbers of data in the partial posteriors. { Note that we keep fixed the total number of posterior evaluations to $E=2000$. This means that as $N$ grows, then $T$ decreases (e.g., in standard LAIS we have $E=2NT$);} {\bf (a)} with $K_n=5$; {\bf (b)} with $K_n=10$. }
	\label{fig:mean errors}
\end{figure}

\noindent
In Figure \ref{fig:mean errors}, we show the obtained results of this a second experiment. In both figures (a)-(b), we see the behavior of the MSE as $ N $ grows {(and also $T$ decreases, since we keep $E=2000$ constant).} The solid line corresponds to the standard LAIS implementation where we use all the data available for the computation of the likelihood in the upper layer. The dashed lines show the behavior of the errors when partial posteriors are considered in the upper layer. The left side shows the case $K_n=5$ for all $n$, while, on the right side, we show $K_n=10$ for all $n$. In both graphics, it can be seen that PLAIS and PA-RLAIS outperform the results of standard LAIS, for the values of $ N $ considered. 
Hence, in this simple example, using partial posteriors improves the performance of the algorithms.
For all methods, the error tends to grow after certain optimal $ N $ { (recall that $T$ is also varying in this figure)}. However, the methods that use partial posteriors show better performance, as compared to standard LAIS, when $N$ increases, that is, when there is more number of shorter chains. 
This can be due to the fact that the partial posteriors are wider, and hence easier to explore in a small number of iterations. 
Also in both cases, the errors of PLAIS and PA-RLAIS are rather similar, although, as exptected, PLAIS outperforms PA-RLAIS. 

\subsection{HMC-LAIS vs HMC algorithms}
\label{sec_second_exp}

For the next experiment, we consider $\post(\x)$ which consists of an equally-weighted mixture of two Gaussian pdfs.
The Gaussians pdfs are located at $[0,0]^\top$ and $[-4,4]^\top$, respectively. The covariance matrix of both is $\bm{\Sigma}=[4,3; 3,4]$.
Here, it is straightforward  to calculate the true values for the quantities of interest: the expected value is $[-2,2]^\top$, the variances are $[8,8]$ and the covariance is $-1$. 
In this simple example, we aim to test the performances  of HMC-LAIS algorithms in estimating these quantities. The goal is to compare their performances against only using HMC algorithms. The error measure we employ is the averaged Mean Squared Error (MSE) in estimating those quantities: expected value of $\post(\x)$ (2 quantities), and covariance matrix of $\post(\x)$ (3 quantities).

The budget is $E=2400$ target evaluations. We consider HMC algorithms with kinetic energy using a Gaussian distribution with covariance matrix equal to $2{\bf I}$, and test the following values for step length and path length \{(0.25,1),(0.5,1),(1,3),(1,5)\}. 
In the lower layer, we also consider Gaussian proposals with covariance matrix equal to $2{\bf I}$. Here, we compare the performance of three deterministic-mixture weighting schemes: spatial, temporal and complete. 

\begin{figure}[h!]
	\centering 
	\centerline{
		\subfigure[HMC parameters (0.25,1)]{\includegraphics[width=7cm]{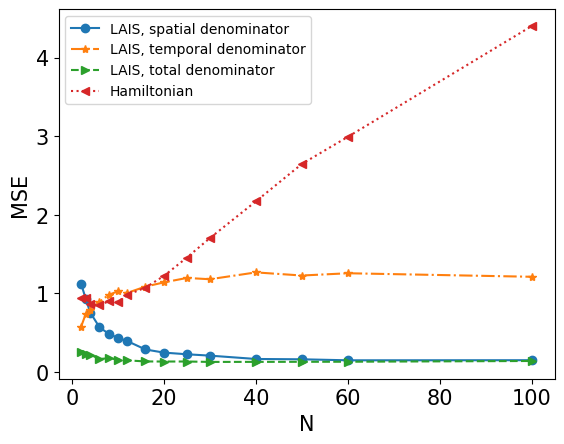}}
		\hspace{0.7cm}
		\subfigure[HMC parameters (0.5,1)]{\includegraphics[width=7cm]{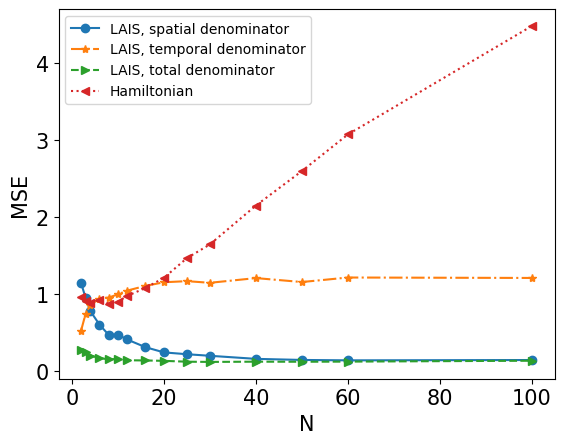}}
	}
	\centerline{
		\subfigure[HMC parameters (1,3)]{\includegraphics[width=7cm]{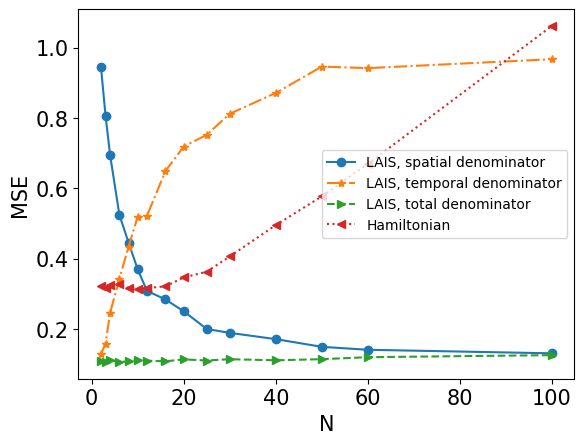}}
		\hspace{0.7cm}
		\subfigure[HMC parameters (1,5)]{\includegraphics[width=7cm]{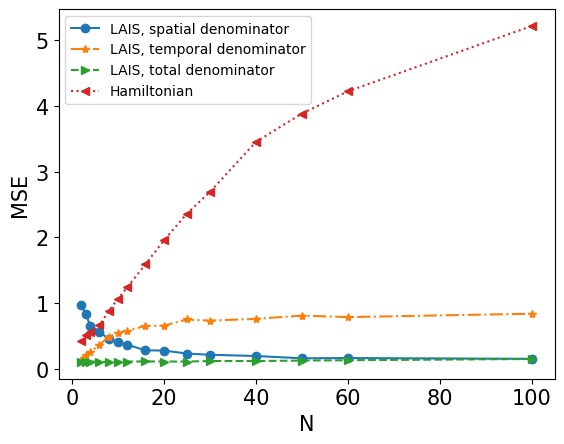}}
	}
	\caption{{{\bf(Fig. of Section \ref{sec_second_exp})}}  MSE in estimation obtained by HMC-LAIS and HMC versus $N$, with the same number of evaluations of the posterior $E=2400$ (hence, the HMC chains have twice the length of the HMC chains used in the upper layer of HMC-LAIS). Each figure corresponds to a different choice of step and path lengths in the HMC algorithms.}
	\label{fig_HMC}
\end{figure}

\begin{figure}[h!]
	\centering 
	\centerline{
		\subfigure[$N=4$]{\includegraphics[width=5.5cm]{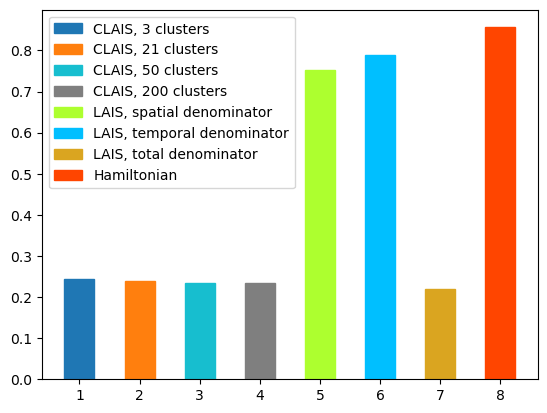}}
		\hspace{0.7cm}
		\subfigure[$N=10$]{\includegraphics[width=5.5cm]{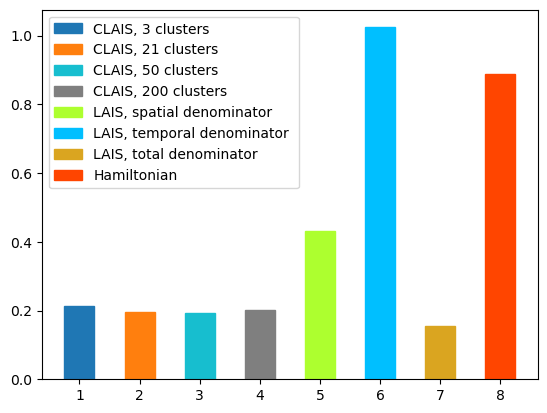}}
	}
	\centerline{
		\subfigure[$N=50$]{\includegraphics[width=5.5cm]{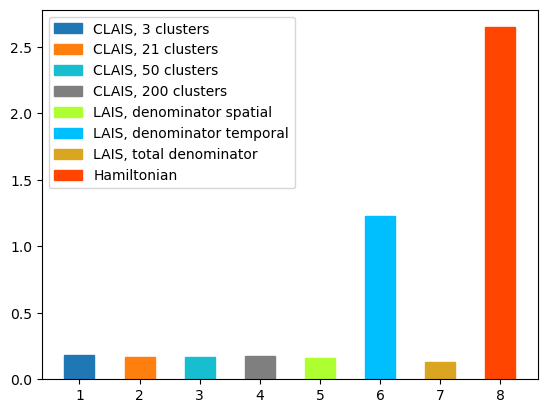}}
		\hspace{0.7cm}
		\subfigure[$N=100$]{\includegraphics[width=5.5cm]{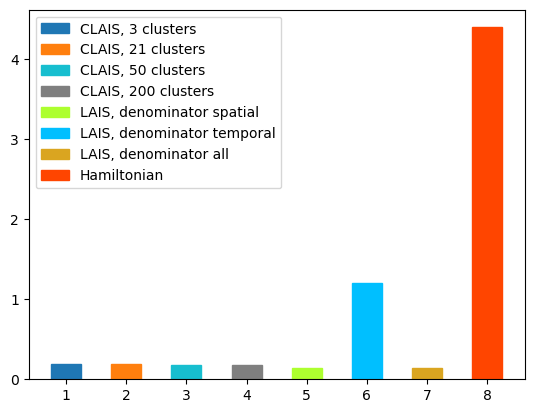}}
	}
	\caption{{{\bf(Fig. of Section \ref{sec_second_exp})}} MSE of CLAIS with different values of $B\in\{3,21,50,200\}$, compared with LAIS with different denominators and parallel HMC chains (with twice lengths with respect to the LAIS schemes, in order to have the same number of posterior evaluations $E=2400$ for all methods). 
}
	\label{fig_con_compression}
\end{figure}

\begin{figure}[h!]
	\centering 
	\centerline{
		\includegraphics[width=7cm]{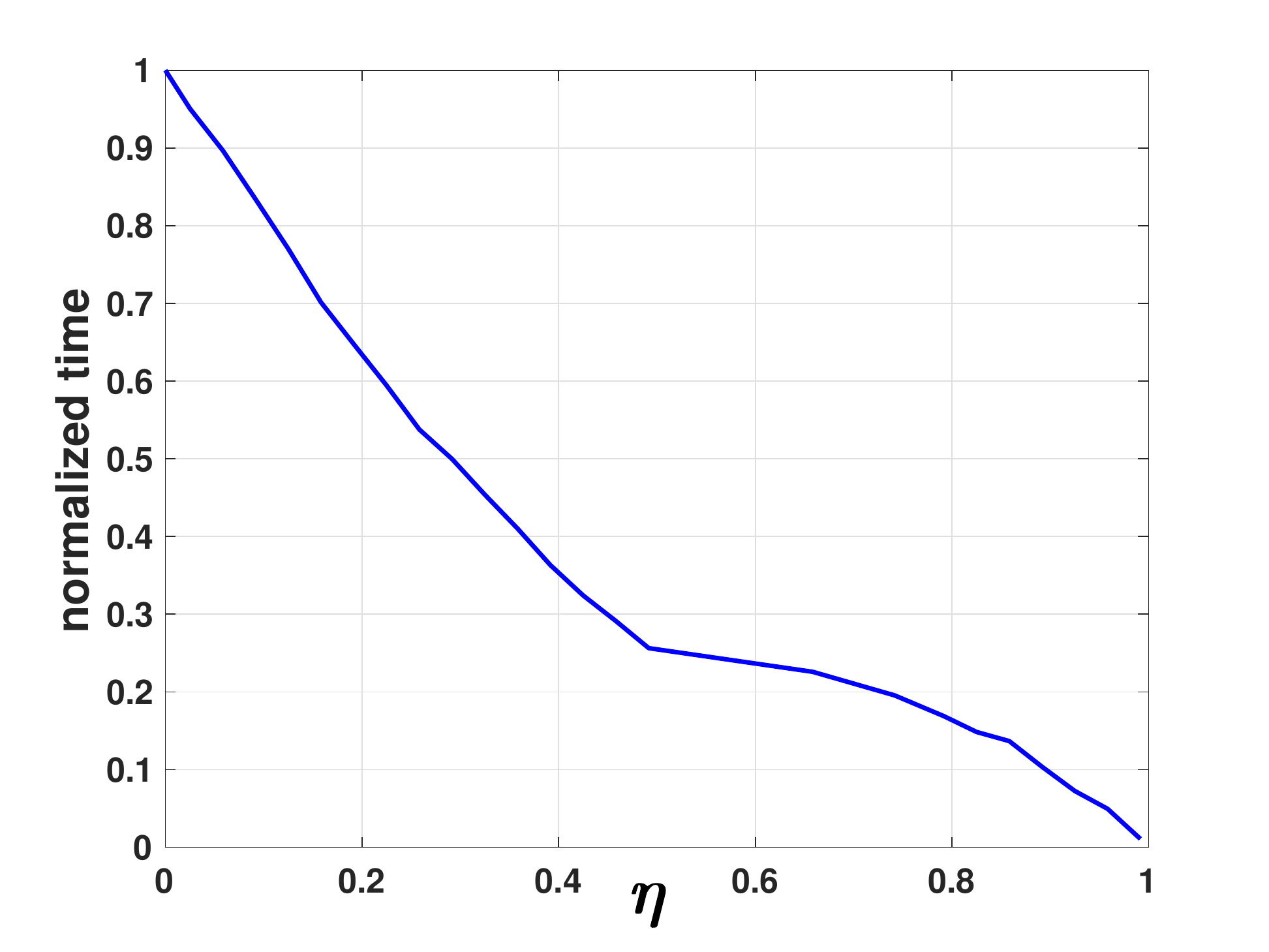}
	}
	\caption{{{\bf(Fig. of Section \ref{sec_second_exp})}} Normalized computational time versus compression level $\eta$, where $\eta=1-\frac{B}{NT}$, and $B$ is the number of clusters.}
	\label{fig_time}
\end{figure}

For setting the number of chains, $ N $, and the number of iterations, $ T $, we follow the same rules as for the previous experiment. We kept constant the 
product $ NT = \frac{E}{2} = 1200 $ and vary $ N $ within \{2,3,4,6,8,10,12,16,20,25,30,40,50,60,100\}.
For a fair comparison, when we only consider HMC algorithms, the $N$ chains were run for $2T$ iterations each (i.e. twice number of iterations than the HMC algorithms in the upper layer of the HMC-LAIS algorithms), so that the final number of target evaluations is $2NT=E=2400$. 
The initial mean vectors were chosen uniformly within the square $ [-10,10]^2 $. 
The results were averaged over 500 independent simulations.
\newline
\newline
In Figure \ref{fig_HMC}, we show the MSE of the HMC and HMC-LAIS algorithms, with three weighting schemes, as a function of $N$.
Recall that, for every $N$, the HMC algorithms were run for twice number of iterations, i.e., they were run for $2T$ iterations, in order to have the same number of target evaluations.
Each figure corresponds to a different choice of step and path lengths in the HMC algorithms. 
\newline
{\bf First main observation.}  We can observe that the LAIS schemes (except some few specific cases) always outperform the HMC algorithms.
\newline
{\bf Second main observation.} It is important to remark the excellent and robust performance provided by HMC-LAIS with the {\it complete} denominator, regardless the parameters of HMC chains  (in the upper layer) used and the number of chains $N$. In fact, HMC-LAIS algorithm with complete denominator clearly outperforms the rest of techniques, providing the smallest error and remaining constant for all $N$ and all HMC parameters. 
\newline
{\bf Other considerations.} The error of HMC is smallest when $N$ is close to the minimum (i.e. when the chains are longer), and gets worse as $N$ increases since, consequently, the chains become shorter and cannot explore properly the two modes.
Interestingly, even in the best scenario, the results show that the error of HMC is always greater than the one provided by HMC-LAIS algorithms with temporal and complete denominators. 
Namely, even when HMC works best, it is better to run it for half number of iterations and then use it within the LAIS framework with a temporal or complete denominator. 
\newline
{\bf Spatial vs Temporal.} The performance of the temporal and spatial denominators behave in an opposite manner.
As expected, the error corresponding to the spatial denominator is worse when $N$ is small. In fact, the greatest error is achieved always when $N$ is minimum. As $N$ increases, the performance greatly improves. It rapidly beats HMC and its performance matches that of the complete weighting scheme for large $N$.
Conversely, in the temporal denominator, the best results are always achieved when $N$ is minimum, since in this case,  the chain length $T$  is maximum. As $N$ increases, the performance of the  temporal denominator worsens, but in a slower fashion than the corresponding error of the HMC algorithms. 
\newline
In this experiment, the spatial denominator seems to outperform the temporal denominator for more values of $N$. 
This means that the mixture of spatial proposals is usually better than the mixture of temporal proposals. For some value $N_*$, both weighting schemes provide the same results. Only for values $N\leq N_*$, the temporal denominator is better than the spatial denominator. Namely, if $T$ is not sufficiently big ($T \geq \frac{E}{2N_*}$), the temporal denominator does not pay off, as compared to the spatial denominator.
In fact, for $N>50$, the spatial denominator can be considered as a compressed version of the complete denominator, i.e., it provides almost the same performance but with a smaller number of components (recall that the complete denominator has $\frac{E}{2}=1200$ mixture components).
\newline
{\bf Compressed schemes.} We have also tested the performance of compressed LAIS (CLAIS), where a compression technique is applied to the $NT$ proposals from the upper layer (see Sect. \ref{sec_compression}). Here, we have run a clustering algorithm with $B\in\{3,21,50,200\}$ clusters to obtain the compressed denominators. In Figure \ref{fig_con_compression}, we show the error of these schemes against the three previous weighting schemes and HMC. With the proposed compression scheme, we see that the performance is very close to that of the complete denominator and it is insensitive to the choice of number of clusters and $N$. 
For moderately low $N$, CLAIS outperforms LAIS with spatial denominator. 
However, as $N$ increases, the spatial denominator matches the performance of CLAIS, i.e., the spatial denominator is also a very efficient way of compressing the $NT$ proposals as discussed above.  
Finally, in Figure \ref{fig_time} we display the computation time of CLAIS versus the compression level $\eta$, which is $\eta=0$ when there is no compression at all ($B=NT$, i.e. the maximum number of clusters), and $\eta=1-\frac{1}{NT}$ when we have $B=1$ clusters. 
%
%
%
%
%

\newpage
\pagebreak
\cleardoublepage

{  
\subsection{High-dimensional experiment}\label{HighDimEx}
\label{sec:SIMU_HD}

In order to be able to compare different scheme  in a high-dimensional sampling problem, we need to know the groundtruth. For this reason, we assume again a mixture of Gaussians as target pdf, i.e.,
\begin{equation}
\bar{\pi}({\bf x})  = \frac{1}{3}\sum_{k=1}^{3} \mathcal{N}({\bf x}; {\bm \nu}_k, \chi_k^2 {\bf I}_{D_X}), \quad {\bf x}\in \mathbb{R}^{D_X},
\label{eq:hdMixture}
\end{equation}  
where ${\bm \nu}_k=[\nu_{k,1}, \ldots,\nu_{k,D_X}]^{\top}$, for $k \in \{1,2,3\}$, with ${\bf I}_{D_X}$ being the $D_X \times D_X$ identity matrix and $D_X$ is the dimension of the space.
In this section, we vary the dimension of the state space in Eq. \eqref{eq:hdMixture} considering $2\leq D_X \leq 50$.
Moreover, we set $\nu_{1,j}=-5$, $\nu_{2,j}=6$, $\nu_{3,j}=3$ for all $j = 1,...,D_X$, and $\chi_k=8$ for all $k \in \{1,2,3\}$.
Note that the expected value of the target ${\pi}({\bf x})$ is then $E[{X_j}]=\frac{4}{3}$ for $j=1,\ldots,D_X$. In order to study the performance of different  Monte Carlo methods, we consider the problem of approximating  this expected value  $E[{X_j}]=\frac{4}{3}$. We apply HMC-LAIS considering $N=100$ parallel chains of HMC in the upper layer, each chain with different parameters. The HMC chains require the selection of following parameters: a positive integer number of ``leap-frog steps'' $B$, a positive number for the step size $\zeta$ and the covariance matrix of the Gaussian kinetic energy $\lambda^2 {\bf I}_{D_X}$ (where we set $\lambda=10$.) We select the two first parameters both randomly for each chain and at each run, $B$ uniformly between 1 and 7 (it must be an integer), and $\zeta \in \mathcal{U}([0.01,0.7])$.  
 The proposal pdfs used in the lower layer, $q_{n,t}(\x|{\bm \mu}_{n,t},{\bf C}_{n})$ are Gaussian pdfs with covariance matrices ${\bf C}_{n}=\sigma^2 {\bf I}_{D_X}$ again with $\sigma=10$. We also draw $M>1$ more than one samples from each proposal in the upper layer. More precisely, we set  $M=19$ and the length of the chains $T=100$ because, since $N=100$, we have a total number of target evaluations of $E=(M+1)NT=2\cdot 10^5$.
\newline
 We compare HMC-LAIS  with different benchmark schemes:  {\bf (a)} the standard PMC scheme \cite{Cappe04}, {\bf (b)} $N$ parallel independent MH chains (Par-MH), {\bf (c)} and a Sequential Monte Carlo (SMC) scheme \cite{Moral06}.  For a fair comparison, all the mentioned algorithms have been implemented in such a way that the number of total evaluations of the target is $E=2\cdot 10^5$ as in HMC-LAIS. Moreover, all the proposal pdfs involved in the experiments are Gaussians, with the same covariance matrices for all the techniques. The initial mean vectors in all techniques are selected randomly and independently as ${\bm \mu}_{n,0} \sim \mathcal{U}([-6\times 6]^{D_X})$ for $n=1,\ldots, N$.
%
\newline
 The results are averaged over  $10^3$  independent runs. Figure \ref{fig:HMC} shows (in log-scale) the MSE in the estimation of $E[{\bf X}]$ as a function of the dimension $D_X$ of the support space. We remark that we have kept fixed the number of total evaluations of the target $E=2\cdot 10^5$ for all the techniques. As expected, the performance of all the methods deteriorates as the dimension of the problem, $D_X$ increases, since we maintain fixed the computational cost $E=2\cdot 10^5$. HMC-LAIS always provides the best results, i.e., obtaining the lower MSE values.  

\begin{figure*}[!h]
%
  \centering
  \centerline{
  \includegraphics[width=10cm]{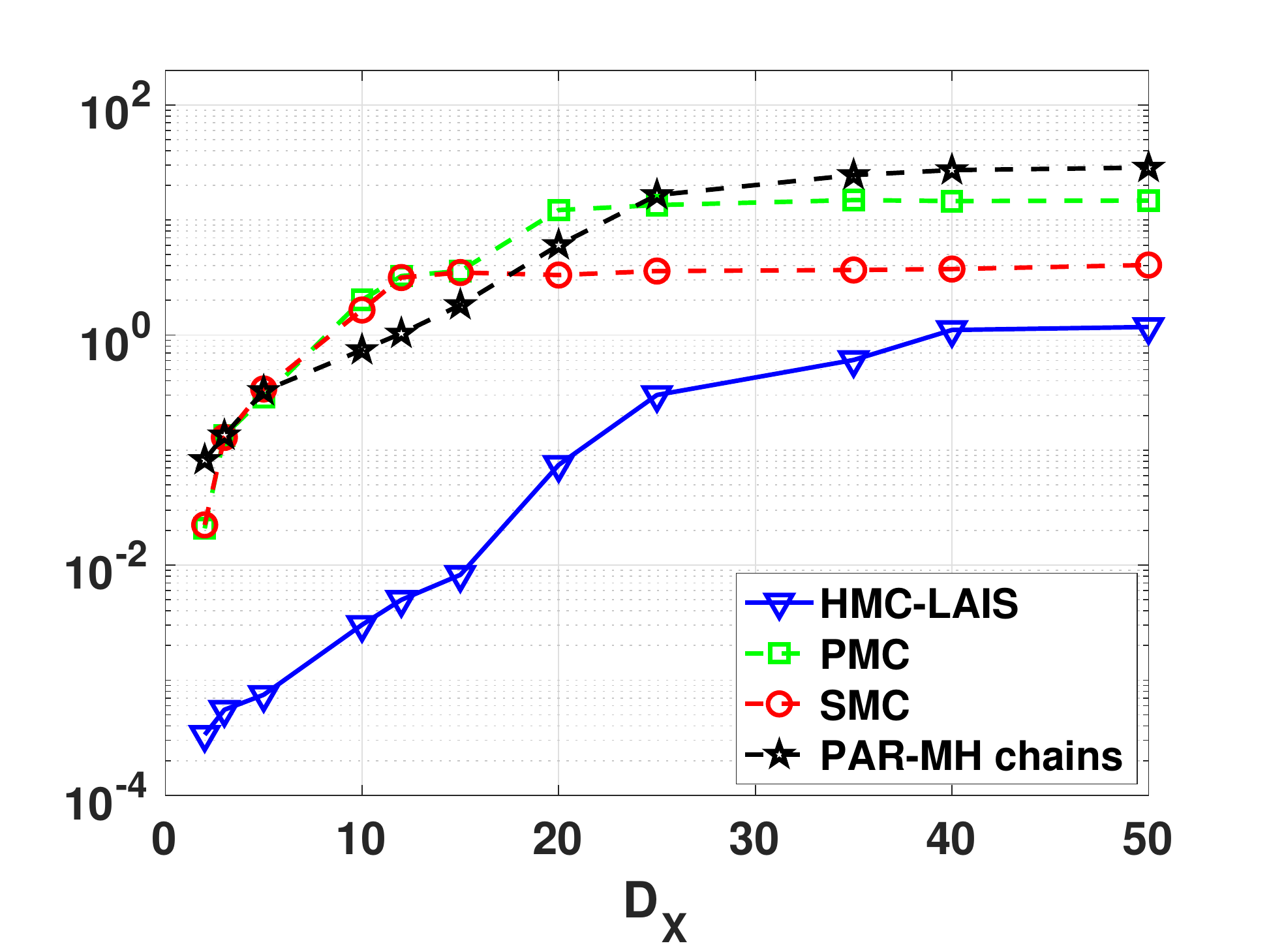}
  }
  \caption{{{\bf(Fig. of Section \ref{sec:SIMU_HD})} MSE (in log-scale) versus the dimension of the space $D_X$,  obtained by the different samplers, with the same total number of target evaluations $E=2\cdot 10^5$. Namely, we keep fixed the computational cost, that in HMC-LAIS means keeping fixed the parameters $N=100$, $M=19$ and $T=100$ (for all $D_X$). }}
\label{fig:HMC}
\end{figure*}

}

{
\subsection{Parameter estimation in a chaotic system}\label{ChaoticSystem}

In this section, we show that the use of Gibbs-LAIS cab be useful in complex inference scenarios where sophisticated MCMC techniques seem to fail \cite{Perretti13,Perretti13b}.
We consider the  estimation problem of parameters in a chaotic system, which is considered a very challenging framework in the literature \cite{Perretti13,Perretti13b,Hartig13}.  This is due to the very tight and sharp posteriors induced by this model. As an example, see as the conditional posterior densities in Figure \ref{fig:Perretti}. The density in Figure \ref{fig:Perretti}(c)  is extremely tight (resembling a delta function), even  sophisticated adaptive Monte Carlo techniques fail. This type of systems are often utilized for modeling the evolution of population sizes, for instance in ecology \cite{Perretti13,Perretti13b}.
Specifically let us consider a logistic map \cite{Boyarsky97} perturbed by multiplicative noise,
\begin{equation}
	y_{k+1}=R\left[\ y_{k}\left(1-\frac{y_k}{\Omega}\right)\right]\exp(\epsilon_k), \quad  \epsilon_k \sim \mathcal{N}(0,\lambda^2),  \quad k=1,...,K,
\label{LogNoisy}
\end{equation}
starting with $y_1 \sim \mathcal{U}([0,1])$. The parameters $R>0$ and $\Omega>0$ are unknown and object of the inference. Hence, using the notation in this work, we have $\x=[R,\Omega]$.
Let us assume that a sequence $\y=y_{1:K}=[y_1,\ldots, y_K]$ is observed and, for the sake of simplicity, let us consider that the standard deviation $\lambda$ of the noise is known. The corresponding likelihood function is given by
\begin{equation*}
L(\y|\x)=p(y_{1:K}|R,\Omega)=\prod_{k=1}^{K-1} p(y_{k+1}|y_k,R,\Omega),
\end{equation*}
where, denoting $g(y_k,R,\Omega)=R\left[\ y_{k}\left(1-\frac{y_k}{\Omega}\right)\right]$, we have 
\begin{equation*}
	p(y_{k+1}|y_k,R,\Omega)\propto \left|\frac{g(y_k,R,\Omega)}{y_{k+1}}\right| \exp\left(-\frac{\log\left(\frac{y_{k+1}}{g(y_k,R,\Omega)}\right)^2}{2\lambda^2}\right),  \quad \mbox{if $g(y_k,R,\Omega)>0$,}
\end{equation*}
 and $p(y_{k+1}|y_k,R,\Omega)=0$, if $g(y_k,R,\Omega)\leq 0$. We set uniform priors, $R \sim \mathcal{U}([0,10^4])$ and $\Omega \sim \mathcal{U}([0,10^4])$, our goal is computing the mean of the bivariate posterior pdf, $\bar{\pi}(\x|\y)=p(R,\Omega| y_{1:K}) \propto p(y_{1:K}|R,\Omega)$, which represents to the minimum mean square error estimator of the vector parameter $\x=[R,\Omega]$ (computing the MSE obtained by the different techniques).
\newline
We have generated artificial data $\y=y_{1:K}$, setting $R=3.7$, $\Omega=0.4$ and $K=20$ (i.e., a trajectory of $20$ values). We employ different values of standard deviation $\lambda=\{0.001,0.005,0.01,0.05,0.08,0.1\}$ of the noise in the system \eqref{LogNoisy} of the same order of magnitude considered in \cite{Perretti13}.
We apply a Gibbs-LAIS scheme where, for drawing from the full-conditional pdfs, we apply (within the Gibbs sampler) the so-called FUSS technique proposed in \cite{MARTINOfuss}. For simplicity, we consider a unique Gibbs chain ($N=1$) in the upper layer with length $T=25$ iterations, i.e., ${\bm \mu}_1,...,{\bm \mu}_T$. In the lower layer of Gibbs-LAIS scheme, we consider two-dimensional Gaussian proposals $q(\x|{\bm \mu}_t)=\mathcal{N}(\x|{\bm \mu}_t, \sigma_p^2{\bf I}_2)$ with $ \sigma_p=1$ and ${\bf I}_2$ is the $2\times 2$ identity matrix. We draw one sample from each proposal $q(\x|{\bm \mu}_t)$, hence we have $S=25$ in the lower layer.
Therefore, the total number of posterior evaluations of the  Gibbs-LAIS scheme is $E=25+25=50$.  Since we have only one chain ($N=1$), we use a temporal weighting scheme.  We also apply the corresponding Gibbs-RLAIS  with the same parameters (then $E=25$), and also we perform a Gibbs-RLAIS but increasing the length of the Gibbs sampler to $T=50$ (so that again $E=50$). Finally, we compare the results with an 
 with MH-within-Gibbs approach with a Gaussian random walk proposal ($ \sigma_p=1$  again) for drawing from the full-conditionals, i.e.,   with $T=50$ steps for the Gibbs samplers, in order to have $E=50$ for a fair comparison.  
For the employed MCMC techniques, the initial states of the chains are chosen randomly from $\mathcal{U}([1,5])$ for $R$ and $\mathcal{U}([0.38,1.5])$ for $\Omega$.
%
\newline
The MSE in estimation obtained by the different techniques (averaged over $1000$ independent runs) is given in Table \ref{TablaResPeretti}. The Gibbs-LAIS schemes outperform clearly the MH-within-Gibbs approach. Moreover, Gibbs-RLAIS with $E=25$ obtains very  close results to Gibbs-LAIS, and Gibbs-RLAIS with $E=50$ even outperforms Gibbs-LAIS when $\lambda$ grows. Another remarkable advantage of employing the Gibbs-LAIS schemes is that one could easily approximating the marginal likelihood $Z=p(\y)=p(y_{1:K})$ in this problem, by computing the estimator $\widehat{Z}$ in \eqref{eq:Z_EST}. In this way, we could perform a model selection study. On the other hand, approximating $Z$ by MH-within-Gibbs method is not a straightforward task \cite{llorente2020marginal}.

%
%


\begin{table*}[ht!]
{
\setlength{\tabcolsep}{2pt}
\def\marginwidth{1.5mm}
\caption{MSEs in estimation of $R$ and $\Omega$, obtained by the different compared techniques.}
\label{TablaResPeretti}
\vspace{-0.3cm}
\begin{center}
\footnotesize
\begin{tabu}{|[1pt]l@{\hspace{\marginwidth}}|c@{\hspace{\marginwidth}}|[1pt]c@{\hspace{\marginwidth}}|c@{\hspace{\marginwidth}}|c@{\hspace{\marginwidth}}|c@{\hspace{\marginwidth}}|c@{\hspace{\marginwidth}}|c@{\hspace{\marginwidth}}|[1pt]}
\cline{3-8}
 \multicolumn{2}{c|[1pt]}{} & $\lambda=0.001$ & $\lambda=0.005$  & $\lambda=0.01$ & $\lambda=0.05$  & $\lambda=0.08$ & $\lambda=0.10$      \\
\Xhline{2\arrayrulewidth}
\multirow{2}{*}{Gibbs-LAIS ($E=50$)} &MSE($R$) & 0.0065 & 0.0067  & 0.0085  & 0.0125 & 0.0142 & 0.0681   \\
\cline{2-8}
  &MSE($\Omega$)  & 4.97  $10^{-5}$ &  6.16  $10^{-5}$ &  4.18  $10^{-5}$ & 5.26  $10^{-5}$ & 6.33  $10^{-5}$ & 1.70  $10^{-4}$  \\
\Xhline{2\arrayrulewidth}
\multirow{2}{*}{Gibbs-RLAIS ($E=25$)} &MSE($R$) & 0.0082 & 0.0090  & 0.0089  & 0.0138 & 0.0160 & 0.0752   \\
\cline{2-8}
  &MSE($\Omega$)  & 5.21  $10^{-5}$ &  6.22  $10^{-5}$ &  6.13  $10^{-5}$ & 4.22  $10^{-5}$ & 5.89  $10^{-5}$ & 1.82  $10^{-4}$  \\
\Xhline{2\arrayrulewidth}
\multirow{2}{*}{Gibbs-RLAIS ($E=50$)} &MSE($R$) & 0.0070 & 0.0069  & 0.0078  & 0.0126 & 0.0130 & 0.0547   \\
\cline{2-8}
  &MSE($\Omega$)  & 5.01  $10^{-5}$ &  6.20  $10^{-5}$ &  5.75  $10^{-5}$ & 5.19  $10^{-5}$ & 6.08  $10^{-5}$ & 1.56  $10^{-4}$  \\
\Xhline{2\arrayrulewidth}
 \multirow{2}{*}{MH-within-Gibbs ($E=50$)} &MSE($R$) &  0.6830& 0.7264 & 0.7067  & 1.1631  &  1.3298 &  1.3293 \\
\cline{2-8}
  &MSE($\Omega$)  & 0.0373 & 0.0402 & 0.0423 &0.0399  & 0.0471 &  0.0440   \\
\Xhline{2\arrayrulewidth}
\hline
\end{tabu}
\end{center}
}
\end{table*}

\begin{figure*}[!h]
%
  \centering
  \centerline{
  \subfigure[Fixing $\Omega=4$ (log-domain).]{\includegraphics[width=5.4cm]{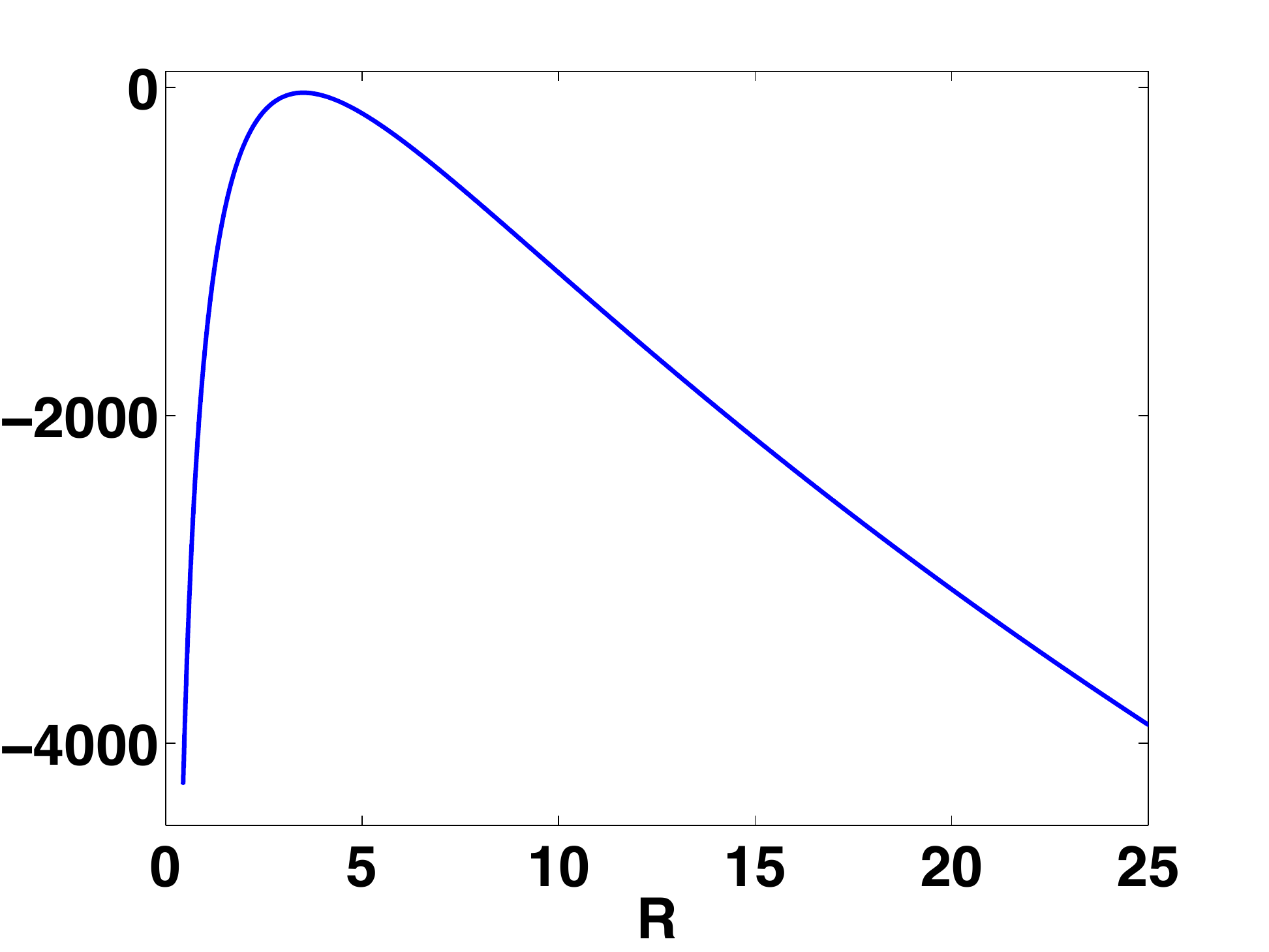}}
  \subfigure[Fixing $R=0.7$ (log-domain).]{\includegraphics[width=5.4cm]{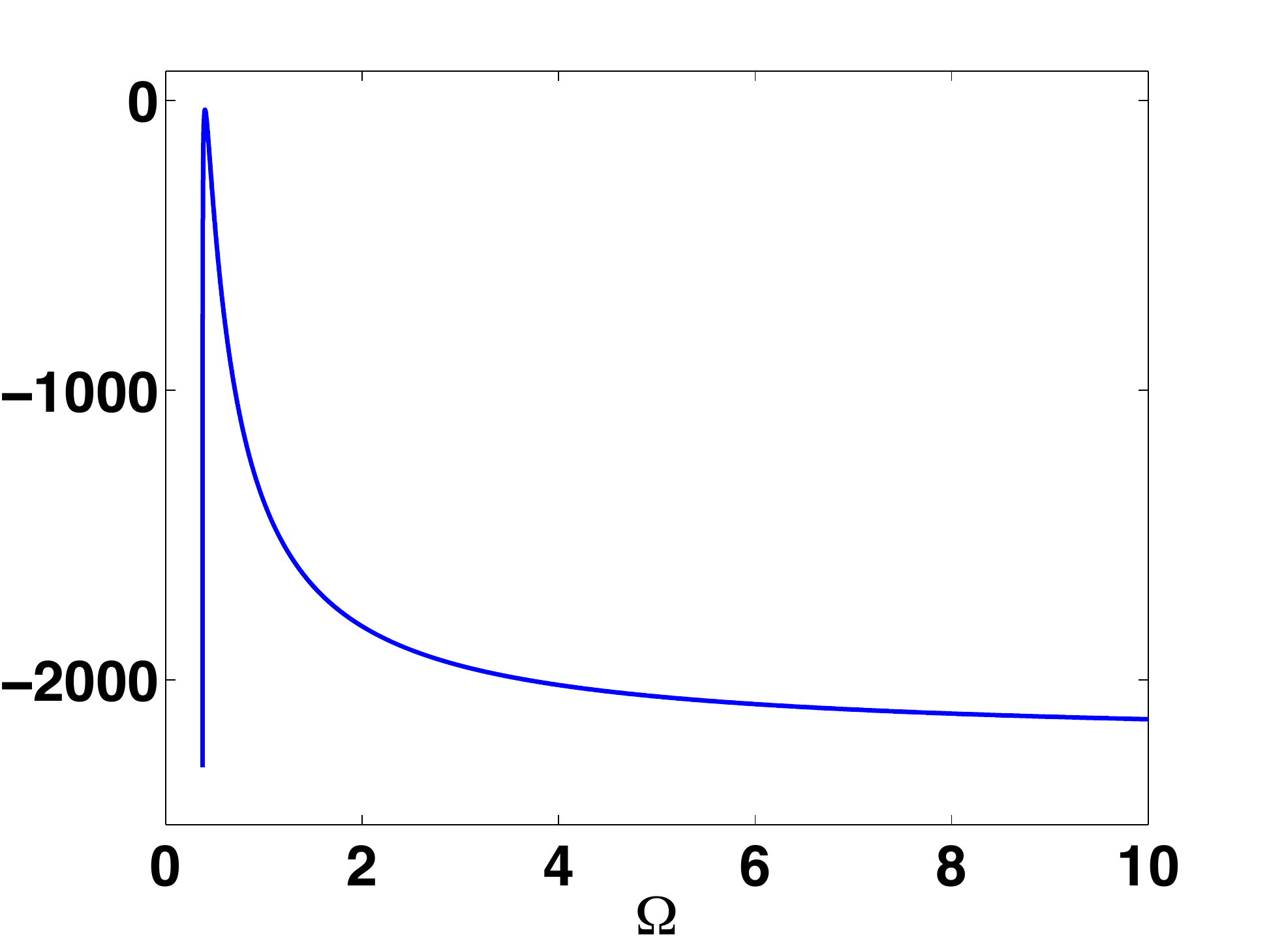}}
  \subfigure[Standard domain.]{\includegraphics[width=5.4cm]{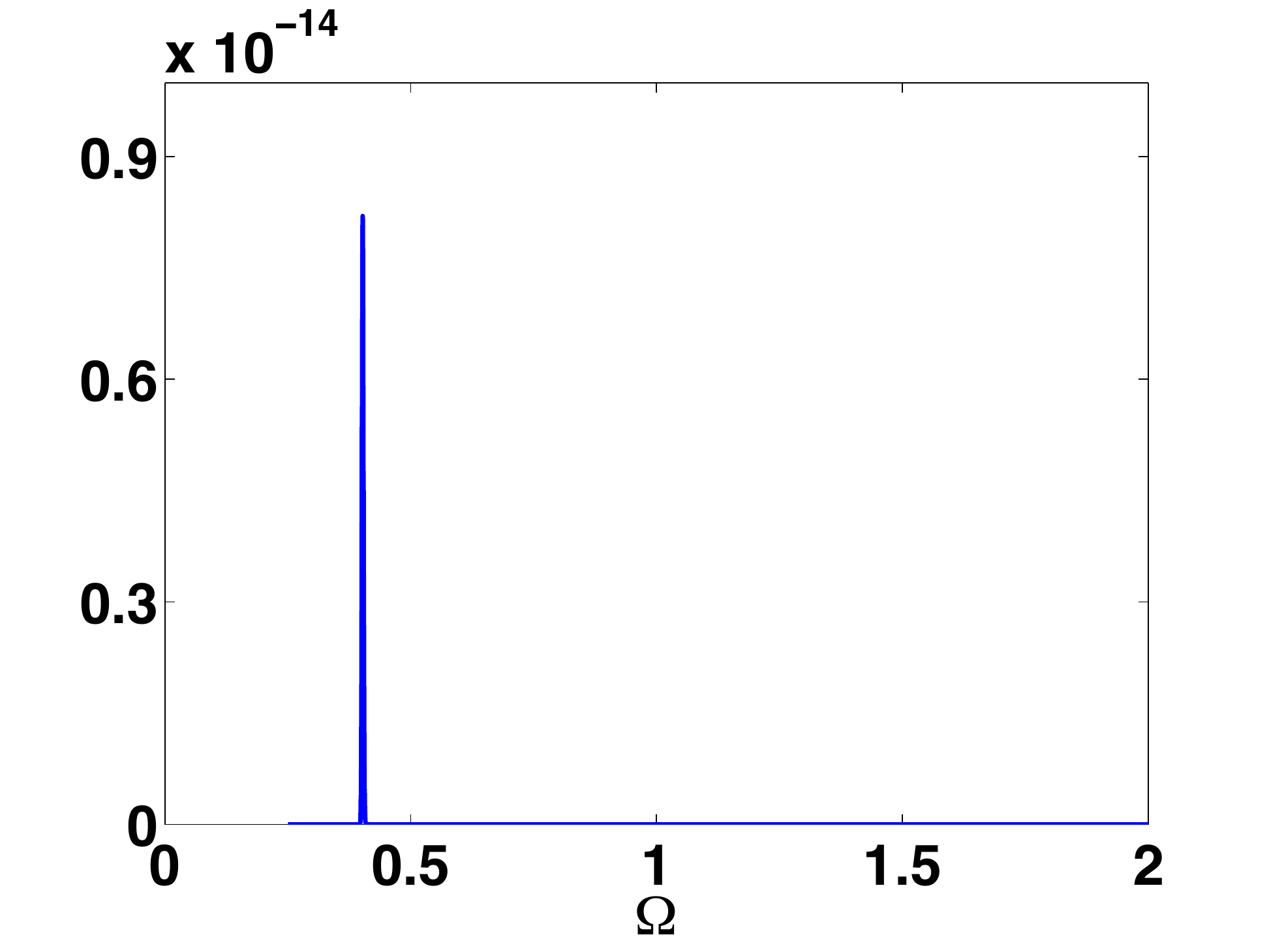}}
  }
%
%
%
\caption{{{{\bf(Fig. of Section \ref{ChaoticSystem})}} {\bf (a)}-{\bf (b)} Examples of conditional densities {\it in log-domain} with $\lambda=0.1$, and considering $K=20$ observations. {\bf (a)} Fixing $\Omega=4$. {\bf (b)} Fixing $R=0.7$. {\bf (c)} The conditional pdf corresponding to Figure (b). Even advanced and adaptive MCMC techniques often fail in drawing samples from this kind of sharp/tight densities.}}
\label{fig:Perretti}
\end{figure*}

}

\newpage
\subsection{Experiment with COVID-19 data}\label{Example covid} 
\begin{figure}[!h]
	\centering
	\centerline{
\includegraphics[width=0.4\textwidth]{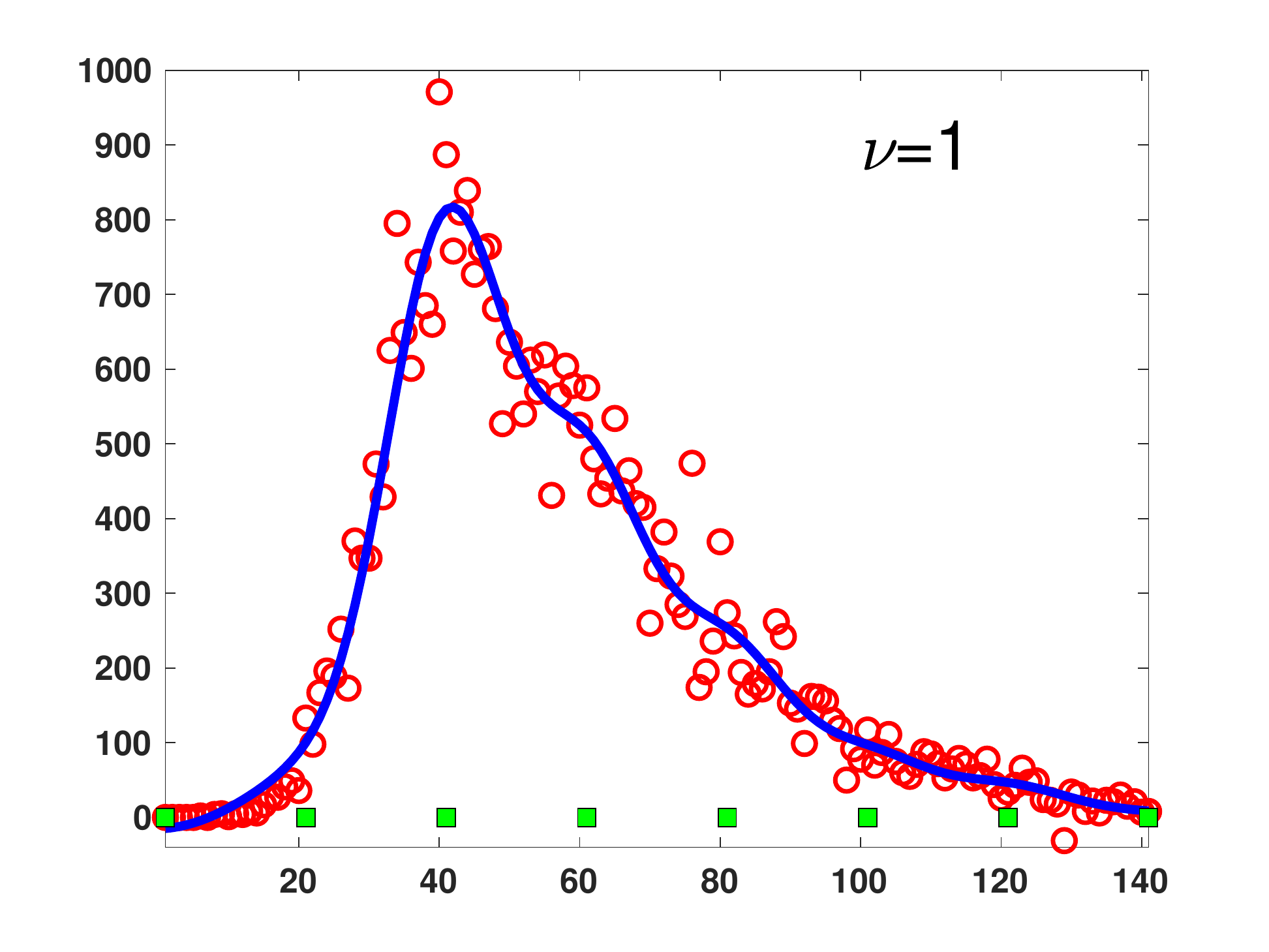}
\includegraphics[width=0.4\textwidth]{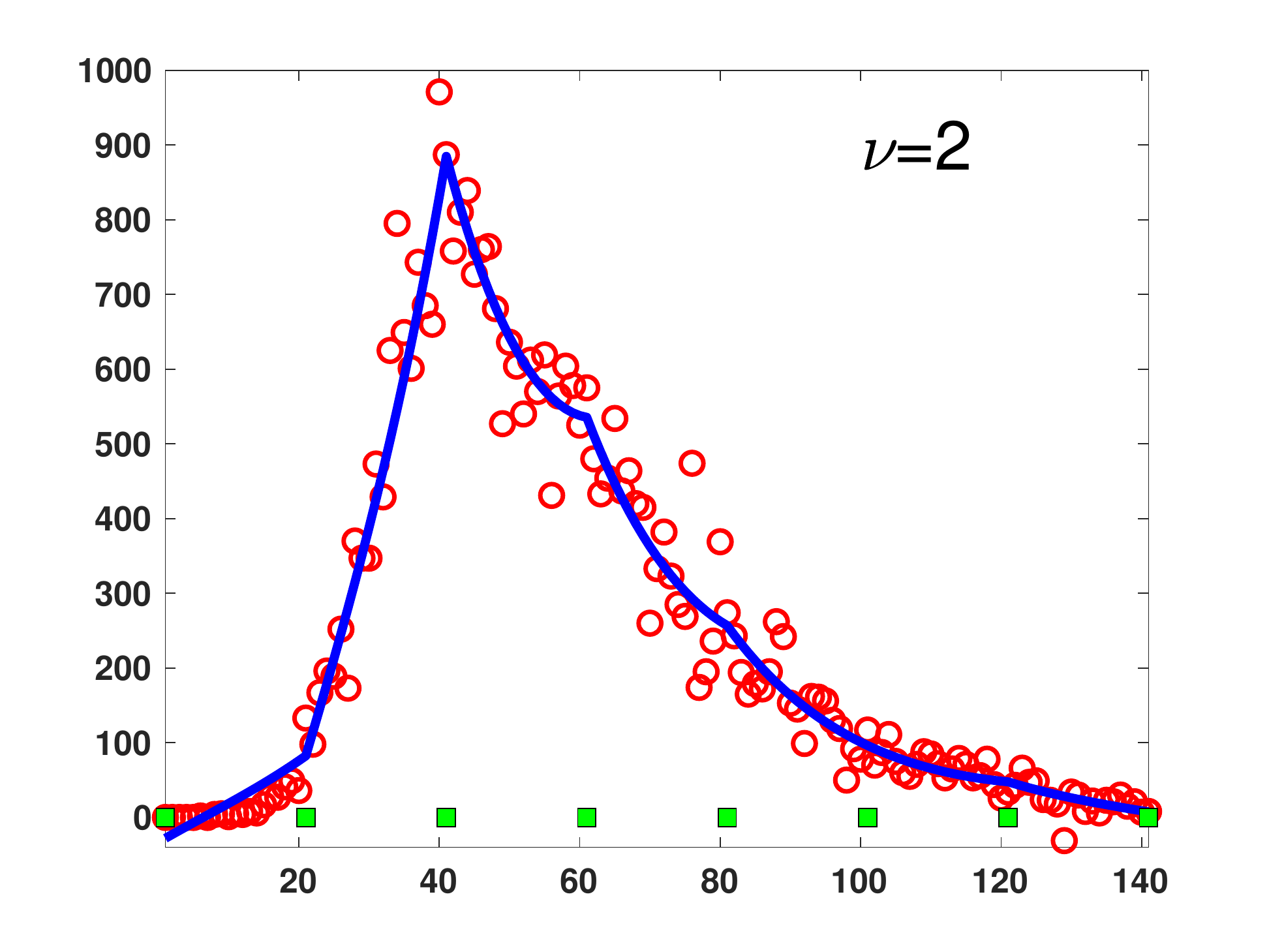}
}
\caption{\footnotesize {{\bf(Fig. of Section \ref{Example covid})}}
	Best fit with $8$ bases with different types of basis, $\nu=1,2$. The circles represent the analyzed data and the squares show the positions of the bases. 
}
\label{Fig_fit}
\end{figure}

We consider the number of daily deaths caused by SAR-CoV-2 in Italy from 18 February 2020 to 6 July 2020 as the dataset.  We denote the values of daily deaths as  $\y = [y_1,\dots,y_{D_Y}]^\top$. Let $t_i$ denote the  $i$-th day, we model each observation as 
	$$
	y_i = f(t_i) + e_i, \quad i=1,\dots,D_Y=140,
	$$
	where $f$ is the function that we aim to approximate and $e_i$'s are  independent Gaussian realizations with zero means and variance $\sigma_e^2$. We consider the approximation of $f$ at some $t$ as a weighted sum of $M$ localized basis functions,
	$$
	{f}(t)=\sum_{m=1}^M { \rho}_m \psi(t|\mu_m,h,\nu),
	$$
	where $\psi(t|\mu_m,h)$ is  $m$-th basis located at $\mu_m$ with bandwidth $h$. Let also be  $\nu$ an  index denoting the type of basis.  We consider $M\in\{1,....,D_Y\}$, then $1\leq M\leq D_Y$. When $M=D_Y$, the model becomes a Relevance Vector Machine (RVM), and the interpolation of all data points (maximum overfitting, with zero fitting error) is possible \cite{bishop2006pattern,rasmussen2003gaussian}.  We study $2$ possible kinds of basis (i.e., $ \nu=1,2$): Gaussian ($\nu=1$),  and Laplacian ($\nu=2$).  After fixing $\nu$ and $M$, we select the locations $\{\mu_m\}_{m=1}^M$ as a uniform grid in the interval  $[1,D_Y]$ (recall that $D_Y=140$).  Hence, by knowing $\nu$ and $M$, the locations $\{\mu_m\}_{m=1}^M$ are given.
\newline	
\newline	
We define the vector of coefficients $\bm{\rho} = [\rho_1,\dots, \rho_M]^\top$.	
 Let also ${\bm \Psi}$ be a $D_Y\times M$ matrix with elements $[\bm{\Psi}]_{i,m} = \psi(t_i|\mu_m,h)$ for $i=1,\dots,D_Y$ and $m=1,\dots,M$. Then, the observation equation in vector form is
	$$
	{\bf y}={\bm \Psi} {\bm \rho}+{\bf e},
	$$ 
where ${\bf e} \sim \mathcal{N}({\bf 0},\sigma_e^2 {\bf I}_{D_Y})$ is a $D_Y \times 1$ vector of noise, where ${\bf I}_{D_Y}$ is the $D_Y\times D_Y$ identity matrix. Therefore, the  likelihood function will be
  $$
  \ell({\bf y}| {\bm \rho},h,\sigma_e, \nu, M)=\mathcal{N}({\bf y}|{\bm \Psi} {\bm \rho},\sigma_e^2 {\bf I}_{D_Y}).
  $$ 
 We assume a Gaussian prior density over the vector of coefficients  ${\bm \rho}$, i.e., $g({\bm \rho}|\lambda) =\mathcal{N}({\bm \rho}|{\bf 0}, {\bm \Sigma}_\rho)$, where $ {\bm \Sigma}_\rho= \lambda {\bf I}_M$ and $\lambda>0$.
Therefore, the complete set of parameters to infer is $\{\bm{\rho}, \nu, M, h, \lambda, \sigma_e\}$.
The conditional posterior of ${\bm \rho}$ given the rest of parameters is also Gaussian, 
	\begin{eqnarray*}
	\post({\bm \rho}|{\bf y},\lambda,h,\sigma_e,\nu, M) = \frac{\ell({\bf y}| {\bm \rho},h,\sigma_e,\nu, M)g({\bm \rho}|\lambda)}{p({\bf y}|\lambda,h,\sigma_e,\nu, M)} 
	= \mathcal{N}({\bm \rho}|{\bm \mu}_{\rho|y}, {\bm \Sigma}_{\rho|y}),   
	\end{eqnarray*}
	and a likelihood marginalized w.r.t. ${\bm \rho}$ is available in closed-form, 
	\begin{equation}
	p({\bf y}|\lambda,h,\sigma_e,\nu, M) =\mathcal{N}({\bf y}|{\bf 0}, {\bm \Psi} {\bm \Sigma}_\rho{\bm \Psi}^{\top}+\sigma_e^2 {\bf I}_{D_Y}).
	\end{equation}
	For further details see \cite{bishop2006pattern,rasmussen2003gaussian}.
	Now, we assume $g_\lambda(\lambda)$, $g_h(h)$, $g_\sigma(\sigma_e)$ are folded-Gaussian priors over $h,\lambda,\sigma_e$,  defined on $\mathbb{R}_+=(0,\infty)$ with location and scale parameters $\{0,100\}$, $\{0,400\}$ and $\{1.5,9\}$, respectively. Then,  we study the following posterior marginalized w.r.t. ${\bm \rho}$ and conditioned to $\mu, M$, 
	$$
	\post(\lambda,h,\sigma_e|\y,\nu, M) = \frac{1}{p(\y|\nu, M)} p(\y|\lambda, h,\sigma_e,\nu, M)g_\lambda(\lambda) g_h(h) g_\sigma(\sigma_e),
	$$  
	Finally, we want to compute the marginal likelihood, i.e.,
	\begin{align}
		 p(\y|\nu, M) = \int_{\mathbb{R}_+^{3}} p(\y|\lambda,h,\sigma_e,\nu, M)g_\lambda(\lambda) g_h(h) g_\sigma(\sigma_e) d\lambda dh d\sigma_e.
	\end{align}
Furthermore,  assuming a uniform probability mass  $p(M=i)=\frac{1}{D_Y}$ as prior over $M$,
we have $p(M|\y,\nu)= \frac{p(\y|\nu, M)p(M)}{p(\y|\nu)} \propto \frac{1}{D_Y}p(\y|\nu, M)$. We can marginalize out $M$ obtaining
	\begin{align}
p(\y|\nu)= \frac{1}{D_Y} \sum_{M=1}^{D_Y}p(\y|\nu, M),   \quad \mbox{ for } \quad \nu=1,2. 
	\end{align}
	Considering also a uniform prior over $\nu$, we can obtain the marginal posterior $
	p(\nu|\y) \propto \frac{1}{2}p(\y|\nu)$.
	\newline
 {\bf Goal.}  Our purpose is: (a) to make inference regarding the parameters of the model $\{\lambda,h,\sigma_e\}$, (b) approximate $Z=p(\y|\nu, M)$, (c) study the posterior $p(M|\y,\nu)$. 
We also study the marginal posterior $p(\nu|\y)$ for $\nu=1,2$. 
  \newline
 {\bf Methods.}  For approximating $p(\y|\nu, M)$, for $M=1,\dots,D_Y$, and $p(\nu|\y)$,  we first apply a  Naive Monte Carlo (NMC) method with $10^4$ samples. We apply also a Gibbs-LAIS scheme with a MH-within-Gibbs sampler in the upper layer.
 More  specifically, we employ an interpolative piecewise constant function as proposal in the MH scheme to draw from the full-conditionals (considering $2$ internal steps)   \cite{MARTINOfuss}.
  Hence, in the upper layer,  we obtain a unique Markov chain ($N=1$) of ${\bm \mu}_t=[\lambda_t,h_t,\sigma_{e,t}]$ for $t=1,\dots,T$.  We set $T=5000$, hence also $5000$ samples drawn in the lower layer and used in estimators. The total number of evaluations of the posterior is  $2T=10^4$ for both, NMC and Gibbs-LAIS schemes.
  \newline
	{\bf Results.} With both methods,  We obtain that MAP estimator of $M$ is $M^*=8$.
	 In Figure \ref{Fig_fit}, we show  the fitting obtained with $M=8$ bases and the parameter estimations provided by the Gibbs-LAIS scheme.	
	Thus, a first conclusion is that the results obtained with models such as RVMs and Gaussian Processes (GPs) (both having $M=140$ \cite{rasmussen2003gaussian,bishop2006pattern}) can be approximated in a very good way with a much more scalable model, as our model here with only  $M=8$  \cite{rasmussen2003gaussian,bishop2006pattern}.  Regarding  the marginal posterior $p(\nu|\y)$, we can observe the results in Table \ref{Locura_ComemeTheDickWithIcreasingStrength}.  With the results provided by both schemes, we should prefer slightly the Laplacian basis. These considerations are reasonable after having a look at Figure \ref{Fig_fit}. 

\begin{table}[!h]
\caption{The approximate marginal posterior $p(\nu|\y)$ with different techniques.} \label{Locura_ComemeTheDickWithIcreasingStrength}
\vspace{-0.4cm}
\begin{center}
	\begin{tabular}{|c|c|c|}
	\hline
		  {\bf Method}  &  $p(\nu=1|\y)$  &  $p(\nu=2|\y)$    \\
		\hline
	         NMC     & 0.4831  &    0.5169  \\
	         Gibbs-LAIS & 0.4930   & 0.5070   \\
		\hline
	\end{tabular}
\end{center}

\end{table}

\section{Conclusions}\label{ConclSect}

We show LAIS is a flexible framework for designing efficient and robust AIS algorithms. 
Furthermore, we have introduced several enhancements in the LAIS framework in order to improve the performance and reduce the overall computational cost.
Specifically, we have proposed that the MCMC algorithms in the upper layer address different partial posteriors (i.e., posteriors of subsets of data) to improve the mixing of the chains due to the data-tempering effect, and at the same time, reducing the costs of the upper layer. 
We have also studied  the use of sophisticated MCMC algorithms, such as HMC and advanced Gibbs techniques, in the upper layer. 
The resulting algorithms inherit the good mixing properties and additionally provide an additional estimator of the marginal likelihood. 
We have discussed different strategies to reduce the cost of the lower layer based on recycling and compression. Specifically, we introduce a novel scheme which recycles all the samples from the upper layer and form the final estimator without further evaluations of the posterior. Moreover, we have shown how to build efficient but cheaper IS weights with a compressed MIS denominator.
Numerous numerical experiments show that the proposed schemes outperform standard applications of LAIS and other benchmark algorithms.

%

\bibliographystyle{plain}
\bibliography{bibliografia}

\begin{appendices}

{
\section{On the choice of  $p({\bm \mu})$ in the upper layer}\label{Choiceofpmu}

\subsection{Theoretical considerations: optimal invariant distribution in upper layer}

Let us consider a hierarchical procedure which mimics the LAIS sample generation approach. For this purpose, we consider a single proposal pdf $q$ in the lower layer defined by the mean $\bm{\mu}\in\mathbb{R}^{D_X}$ and scale matrix ${\bf C}\in\mathbb{R}^{D_X\times D_X}$, so that the proposal can be denoted as $q(\mathbf{x}|\bm{\mu},\mathbf{C})$, and it fulfills  $q(\mathbf{x}|\bm{\mu},\mathbf{C})=
q(\mathbf{x}-\boldsymbol{\mu}  |{\bf 0}, \mathbf{C})$.\footnote{This property is satisfied by relevant distributions such as Gaussian, Student's t and Laplace.}
We assume that the location parameter $\boldsymbol{\mu}$ is  drawn exactly from the density  $p(\boldsymbol{\mu})$.\footnote{This is clearly a simplification since, with MCMC chains, we obtain correlated samples.}
Hence, the simplified LAIS generation procedure is given below:
\begin{enumerate}
\item Draw a possible location parameter $\boldsymbol{\mu}' \sim p(\boldsymbol{\mu})$.
\item Draw $\mathbf{x} \sim q\left(\mathbf{x}  | \boldsymbol{\mu}', \mathbf{C}\right)$.
\end{enumerate}
 Note that $p(\boldsymbol{\mu})$ plays the role of a prior pdf over the location parameter of $q$. The sample $\mathbf{x}$ is distributed according to the following equivalent density,
\begin{eqnarray}\label{equivalentProposal}
\widetilde{q}(\mathbf{x}  | \mathbf{C})=\int_{\mathcal{X}} q(\mathbf{x}  | \boldsymbol{\mu}, \mathbf{C}) p(\boldsymbol{\mu}) d \boldsymbol{\mu} =\int_{\mathcal{X}} q(\mathbf{x}- \boldsymbol{\mu}|{\bf 0}, \mathbf{C}) p(\boldsymbol{\mu}) d \boldsymbol{\mu},
\end{eqnarray}
i.e., $\mathbf{x} \sim \widetilde{q}(\mathbf{x}  | \mathbf{C})$. 
From Eq. \eqref{equivalentProposal} we can deduce the following considerations.
The last expression in \eqref{equivalentProposal} is a convolution integral. Hence, considering the sum of two independent random variables  
	\begin{equation}\label{Sum2RV}
		{\bf X}={\bf Z}+{\bf M},
	\end{equation}
	where ${\bf Z} \sim q(\mathbf{x}  |{\bf 0}, \mathbf{C})$  (with ${\bm \mu}=0$) and ${\bf M}\sim p(\boldsymbol{\mu})$, then ${\bf X}$ is distributed as $\widetilde{q}(\mathbf{x}|\mathbf{C})$ \cite{Robert04}.
\newline
Now, we consider the problem of finding the optimal density $p^{*}(\bm{\mu}|\mathbf{C})$ over the location parameter $\boldsymbol{\mu}$.  
In LAIS, the samples obtained by this procedure are then used in a self-normalized importance estimator. 
The variance of the IS weights is minimized when the proposal  is exactly $ \bar{\pi}(\mathbf{x}|{\bf y}_{\texttt{tot}})$ \cite{Deniz2021ConvAIS,Robert04}.
Therefore, the desirable scenario is to have $\widetilde{q}(\mathbf{x} | \mathbf{C})=\bar{\pi}(\mathbf{x}|{\bf y}_{\texttt{tot}})$.
The optimal pdf depends on the chosen scale parameter $\mathbf{C}$ and since $q(\mathbf{x}  | \boldsymbol{\mu}, \mathbf{C})=q(\mathbf{x}-\boldsymbol{\mu}|{\bf 0}, \mathbf{C})$, as $\boldsymbol{\mu}$ is a location parameter, we can write
\begin{equation}\label{equivalentProposal2}
  \bar{\pi}(\mathbf{x}|{\bf y}_{\texttt{tot}})=\int_{\mathcal{X}} q(\mathbf{x}-\boldsymbol{\mu}|{\bf 0}, \mathbf{C}) p^{*}(\boldsymbol{\mu}  | \mathbf{C}) d \boldsymbol{\mu}.
\end{equation}
 Equation \eqref{equivalentProposal2} above can be rewritten in terms of the characteristic functions: $Q(\boldsymbol{\nu}  | \mathbf{C})=\int q(\mathbf{x}  | {\bf 0}, \mathbf{C}) e^{i \boldsymbol{\nu}^\top \mathbf{x}}d\x$, $P^{*}(\boldsymbol{\nu}  | \mathbf{C})=\int p^{*}(\mathbf{x}  | \mathbf{C}) e^{i \boldsymbol{\nu}^\top \mathbf{x}}d\x$,
and
$\bar{\Pi}(\boldsymbol{\nu})=\int \bar{\pi}(\mathbf{x}|{\bf y}_{\texttt{tot}}) e^{i \boldsymbol{\nu}^\top {\bf x}}d\x$,  where
$\bm{\nu}\in \mathbb{R}^{D_X}$. 
The characteristic function of ${\bf X}$ is the product of characteristic functions of ${\bf Z}$ and ${\bf M}$.
Hence, in some cases, the optimal invariant pdf in the upper layer has the following characteristic function,\footnote{$P^{*}(\boldsymbol{\nu}  | \mathbf{C})$ could not define a pdf. In this case, the optimal  invariant pdf cannot be expressed as in Eq. \eqref{EqPop}.}
\begin{align}\label{EqPop}
	P^{*}(\boldsymbol{\nu}  | \mathbf{C})=\frac{\bar{\Pi}(\boldsymbol{\nu})}{Q(\boldsymbol{\nu}  | \mathbf{C})}.
\end{align}
In a general case, it is not possible to determine analytically the expression of the optimal pdf $p^{*}(\boldsymbol{\mu}  | \mathbf{C})$, and thus, other practical choices must be considered, as discussed below.


\subsection{Practical choices of the invariant distribution in the upper layer}

Here, we discuss some practical selection of $p({\bm \mu})$. First of all,
from Eq. \eqref{Sum2RV}, we can obtain the following relevant considerations for this purpose:
\begin{enumerate}
\item  $\mbox{E}[{\bf X}]=\mbox{E}[{\bf Z}]+\mbox{E}[{\bf M}]=0+\mbox{E}[{\bf M}]$, i.e., the expected value of the equivalent proposal $\widetilde{q}$ is equal to the expected value of the density $p({\bm \mu})$ in the upper layer.
\item $\mbox{Var}[{\bf X}]=\mbox{Var}[{\bf Z}]+\mbox{Var}[{\bf M}]\geq \mbox{Var}[{\bf M}]$, where $\mbox{Var}[\cdot]$ returns the elements in the diagonal of the covariance matrix and, the inequality $\geq$ is applied to each element in the diagonal. Namely, the variances of each component of the equivalent proposal $\widetilde{q}$ are greater or equal to  the variances of each component of the density $p({\bm \mu})$ in the upper layer. 
\end{enumerate}
Thus, the equivalent density $\widetilde{q}(\mathbf{x}|\mathbf{C})$ has the same  expected value and a bigger variance with respect to the density $p({\bm \mu})$. 
\newline
\newline
{\bf Consideration on the optimal pdf $p^*({\bm \mu})$.} Given Eq. \eqref{equivalentProposal2} and the observations above, we can deduce that the optimal pdf $p^*({\bm \mu})$ will have the same mean as the posterior, and it will have lighter tails than the posterior $\bar{\pi}$ (i.e., $p^*$ is more ``concentrated'' than $\bar{\pi}$).
\newline
\newline
{\bf A possible choice of $p({\bm \mu})$ in the upper layer.} 
In practice, we cannot employ the optimal $p^*(\bm{\mu})$.
However, the choice $p({\bm \mu})=\post({\bm \mu}|{\bf y}_{\texttt{tot}})$ provides an equivalent proposal with the same mean as the posterior, but with heavier tails. This is a good property: indeed, it avoids  infinite variance estimators (see example 1 in \cite{llorente2020marginal}) and this is the reason why this choice works well in practice \cite{martino2016layered}.
It can be shown that, in this case, the equivalent proposal is the kernel density estimator of the posterior (for a fixed optimal choice of ${\bf C}$).
However, when there are large amounts of data,  evaluating the posterior can be very costly, so that the upper layer can require too much computational time. 
Furthermore, it is common that $\pi(\x|{\bf y}_{\texttt{tot}})$ is highly concentrated in some regions, so the MCMC algorithms in the upper layer can suffer from bad mixing. Also in this scenario, LAIS is able to provide final consistent  estimators due to the use of weighted samples in the lower layer.

\subsubsection{Standard tempering and anti-tempering}\label{TemperingSect}

One idea for solving the second issue above, i.e., the bad mixing of the MCMC chains when $\pi(\x|{\bf y}_{\texttt{tot}})$ is highly concentrated, is the so-called {\it tempering}. 
 Roughly speaking, tempering is a technique used to artificially change the scale of the target density. 
It is commonly used in order to improve the exploration of the posterior support in optimization, MCMC and IS \cite{earl2005parallel, neal2001annealed}. For instance, taking $p(\bm{\mu}) \propto \pi(\bm{\mu}|{\bf y}_\texttt{tot})^\beta$ with $0<\beta<1$ as the target density can be useful if $\bar{\pi}$ concentrates in a small region that is not easy to discover. The $\beta$ is usually referred to as the (inverse) temperature parameter. 
More generally, a temperature schedule is a sequence of tempered posteriors ending with $\bar{\pi}$. A common choice is the geometric path between prior and posterior
	$\post_{\beta_n}(\x|{\bf y}_\texttt{tot}) \propto \pi(\x|{\bf y}_\texttt{tot})^{\beta_n} g(\x)^{1- \beta_n} = L({\bf y}_\texttt{tot}|\x)^{\beta_n} g(\x)$, 
for a sequence $0=\beta_0<\beta_1<\dots<\beta_N=1$, such $\post_{\beta_0}(\x|{\bf y}_\texttt{tot})=g(\x)$ (i.e., the prior pdf over $\x$) and $\post_{\beta_N}(\x|{\bf y}_\texttt{tot}) = \post(\x|{\bf y}_\texttt{tot})$. 
Note that the tempered posterior has a powered, less informative (i.e., wider) likelihood. 
\newline
Therefore, in order to improve the exploration of the posterior support, one possibility consists in taking $p_n({\bm \mu})=\post_{\beta_n}(\x|{\bf y}_\texttt{tot})$  in the upper layer. 
\newline
{\bf Anti-tempering.} An important point to remark that, in LAIS, we can have $\beta \leq 1$ in order to foster the mixing of the chains, but also we can choose some  $\beta > 1$  since, theoretically, the optimal pdf $p^*({\bm \mu})$  is more ``concentrated'' than the posterior $\bar{\pi}$ (as we have seen above).
\newline
In any case, with a standard tempering strategy (using an auxiliary parameter $\beta$), we  only solve one of the two issues pointed out in the previous section: improving the exploration of the posterior support.  The cost of  evaluating a  tempered posterior $\post_{\beta_n}(\x)$ is the same as the cost of  evaluating the non-tempered posterior $\post$. An alternative to the standard tempering procedure is the so-called {\it data tempering}, which reduces also the evaluation cost.
}

		\section{Hierarchical interpretation of the random walk Metropolis-Hastings (MH) algorithm}\label{HI_MH}
	
\begin{figure*}[!hbt]
		\centering 
		\centerline{
			\subfigure[]{\includegraphics[width=5cm]{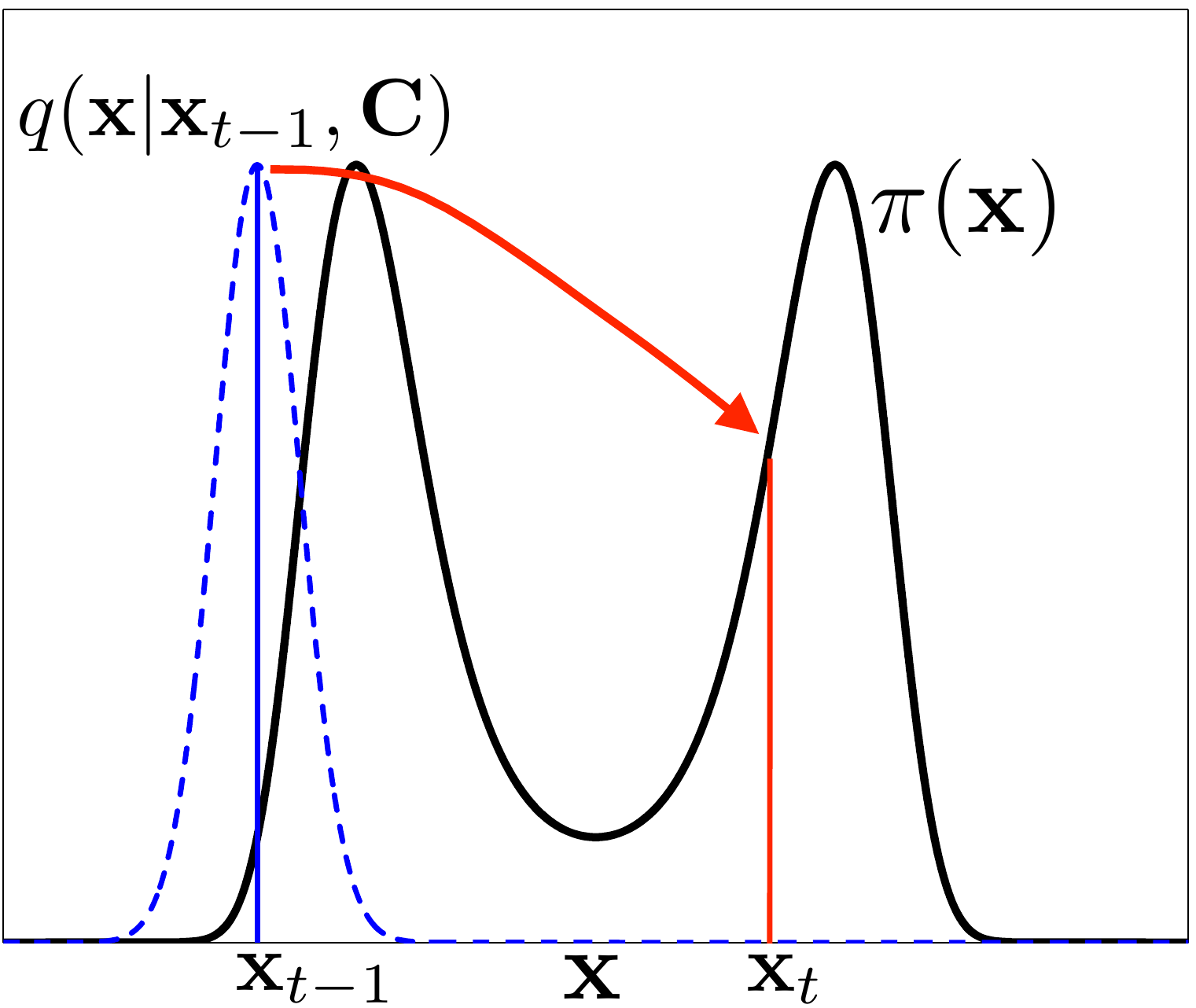}}
			\subfigure[]{\includegraphics[width=5cm]{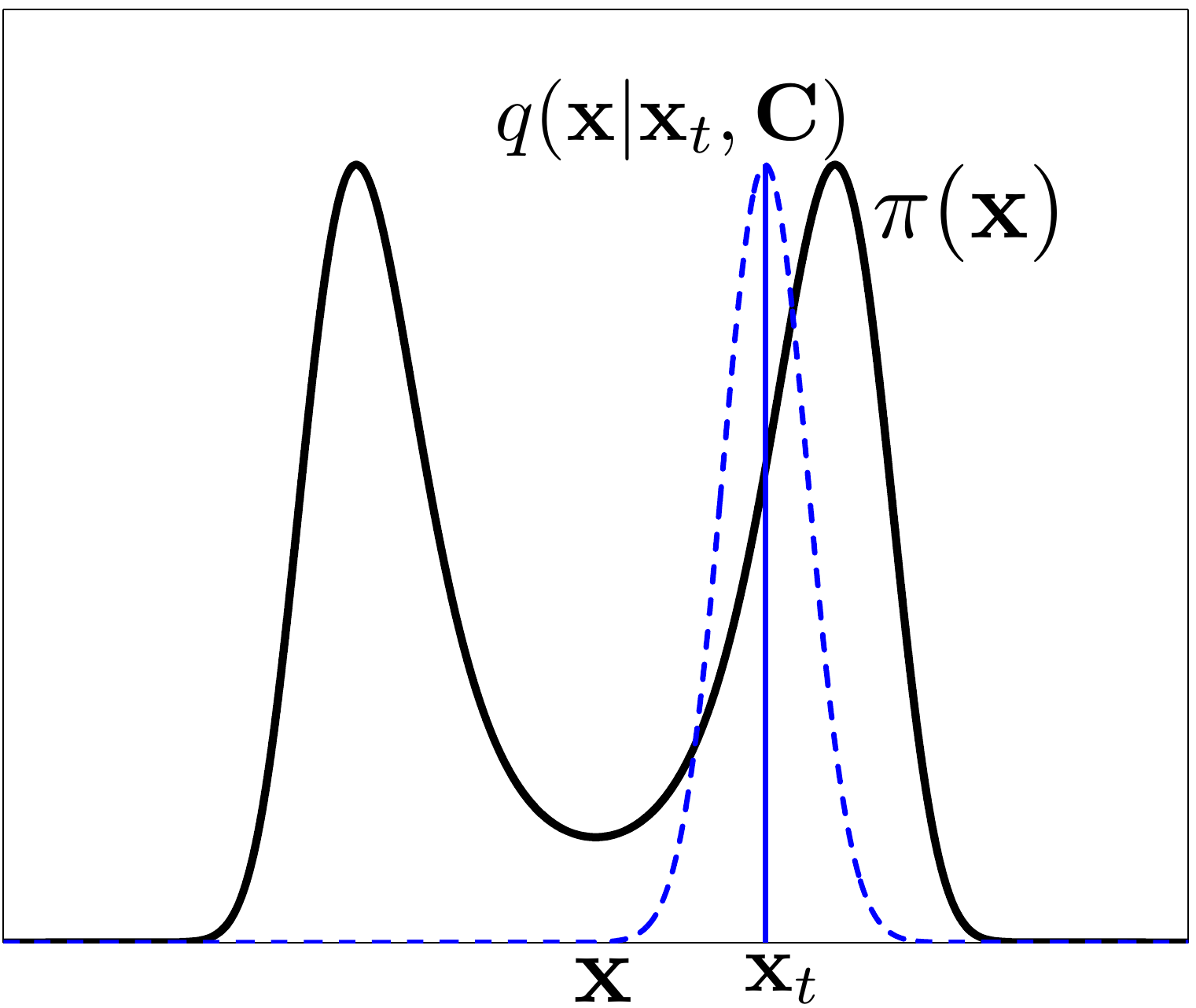}}
			\subfigure[]{\includegraphics[width=5cm]{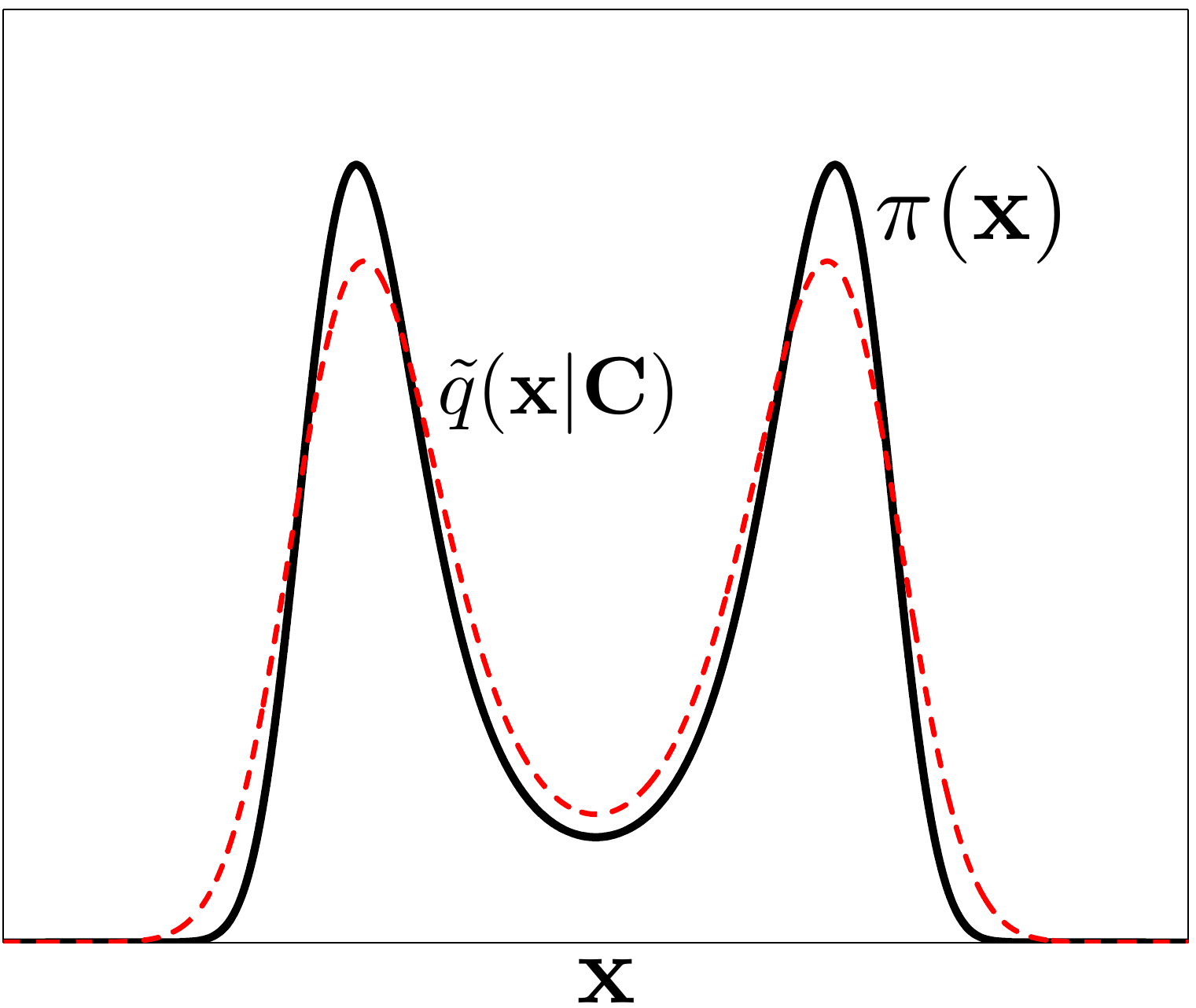}}
		}
		\caption{ Graphical representation of the equivalent proposal of a random-walk proposal in a MH method. A bimodal target pdf $\pi({\bf x})$ is shown in solid line. The proposal densities are depicted in dashed lines. {\bf (a)} A proposal pdf $q({\bf x}|{\bf x}_{t-1},{\bf C})=q({\bf x}-{\bf x}_{t-1}|{\bf 0},{\bf C})$ at the iteration $t-1$, and the next state of the chain ${\bf x}_t$.  {\bf (b)}  The proposal pdf  $q({\bf x}|{\bf x}_{t},{\bf C})=q({\bf x}-{\bf x}_{t}|{\bf 0},{\bf C})$ at the $t$-th iteration.  {\bf (c)} The equivalent independent proposal pdf $\widetilde{q}_{MH}({\bf x}|{\bf C})$ is represented in dashed line.    }
		\label{figEquEX}
	\end{figure*}

	Consider a target density $\pi(\mathbf{x}) \propto \bar{\pi}(\mathbf{x})$ and a random-walk proposal pdf $q\left(\mathbf{x} | \mathbf{x}_{t-1}, \mathbf{C}\right)=q\left(\mathbf{x} - \mathbf{x}_{t-1}| {\bf 0}, \mathbf{C}\right)$, where $\mathbf{x}_{t-1}$ the current state of the chain and $\mathbf{C}$ is a covariance matrix. One transition of the MH algorithm is summarized by
	1. Draw $\mathbf{x}^{\prime}$ from a proposal pdf $q\left(\mathbf{x} | \mathbf{x}_{t-1}, \mathbf{C}\right)$.
	2. Set $\mathbf{x}_{t}=\mathbf{x}^{\prime}$ with probability
	$$
	\alpha=\min \left[1, \frac{\pi\left(\mathbf{x}^{\prime}\right) q\left(\mathbf{x}_{t-1} | \mathbf{x}^{\prime}, \mathbf{C}\right)}{\pi\left(\mathbf{x}_{t-1}\right) q\left(\mathbf{x}^{\prime} | \mathbf{x}_{t-1}, \mathbf{C}\right)}\right]
	$$
	otherwise set $\mathbf{x}_{t}=\mathbf{x}_{t-1}$ (with probability $1-\alpha$ ).
	There are two well-known general classes of proposal pdf: independent proposal $q$ (independent from the current state), and random walk proposal, $q\left(\mathbf{x} | \mathbf{x}_{t-1}, \mathbf{C}\right)$, as we considered above.
	The use of a random walk proposal $q\left(\mathbf{x} - \mathbf{x}_{t-1}|{\bf 0}, \mathbf{C}\right)$ is often preferred due to its explorative behavior, since it relocates the proposal at the current state of the chain at each iteration. See Figure \ref{figEquEX}(a)-(b), for an example. As a consequence, the common wisdom is that this approach is more robust with respect to the choice of the tuning parameters. Below, we provide some further arguments explaining the success of the random walk approach.

We provide a hierarchical interpretation in the same fashion on LAIS. Let us assume a "burn-in" length $T_{b}-1$. Hence, considering an iteration $t \geq T_{b}$, we can assert $\mathbf{x}_{t} \sim \bar{\pi}(\mathbf{x})$.  It implies that the random walk generating process is equivalent, for $t \geq T_{b}$, to the following hierarchical procedure: (a) draw a location parameter $\boldsymbol{\mu}^{\prime}$ from $\bar{\pi}(\boldsymbol{\mu})$, (b) draw $\mathbf{x}^{\prime}$ from $q\left(\mathbf{x} | \boldsymbol{\mu}^{\prime}, \mathbf{C}\right)$.
	Therefore, for $t \geq T_{b}$, the probability of proposing a new sample (i.e., the equivalent proposal) can be written as
	\begin{align}\label{superEq}
	\widetilde{q}_{MH}(\mathbf{x} | \mathbf{C}) &=\int_{\mathcal{X}} q\left(\mathbf{x} | \mathbf{x}_{t-1}, \mathbf{C}\right) \bar{\pi}\left(\mathbf{x}_{t-1}\right) d \mathbf{x}_{t-1}, \nonumber \\
	&=\int_{\mathcal{X}} q\left(\mathbf{x}-\mathbf{x}_{t-1} |{\bf 0}, \mathbf{C}\right) \bar{\pi}\left(\mathbf{x}_{t-1}\right) d \mathbf{x}_{t-1}, \quad \text { for } t \geq T_{b},
	\end{align}
	since $\mathbf{x}_{t-1} \sim \bar{\pi}\left(\mathbf{x}_{t-1}\right)$ after a burn-in period, $t \geq T_{b}$, and $\mathbf{x}_{t-1}$ represents the location parameter of $q$. The function $\widetilde{q}_{MH}(\mathbf{x} | \mathbf{C})$ is an equivalent independent proposal pdf corresponding to a random walk generating process within an MCMC method (after the "burn-in" period).  See Figure \ref{figEquEX}(c) for an example of $\widetilde{q}_{MH}$.
	\newline
	Clearly, this interpretation has no direct implications for practical purposes, since we are not able to draw directly form the target $\bar{\pi}$. However, it is useful for clarifying the main advantage of the random walk approach, i.e., that the equivalent proposal $\widetilde{q}_{MH}$ is a better choice than an independent proposal roughly tuned by the user with non-optimal parameters. In fact, as an example, Eq. \eqref{superEq} ensures that the equivalent proposal $\widetilde{q}_{MH}(\mathbf{x} | \mathbf{C}) $ has a fatter tails than the target $\bar{\pi}$.
	Indeed, the random walk generating procedure includes indirectly certain information about the target: denoting ${\bf X} \sim \tilde{q}_{MH}({\bf x} | {\bf C})$, ${\bf Z}\sim q\left(\mathbf{x}|{\bf 0}, \mathbf{C}\right)$ and ${\bf M}\sim \bar{\pi}({\bf x})$, we have 
	$$
	E[\mathbf{X}]=E[\mathbf{M}] , \quad \mathbf{\Sigma}_{X}=\mathbf{C}+\mathbf{\Sigma}_{M},
	$$
	where  $E[\mathbf{M}]$ and $\mathbf{\Sigma}_{M}$ are the mean and covariance matrix of the target pdf  $\bar{\pi}({\bf x})$. 

\end{appendices}

\end{document}